\documentclass[%
 reprint,
 superscriptaddress,
 amsmath,amssymb,
 aps,
pra,
 floatfix,
]{revtex4-1}
\pdfoutput=1

\usepackage{graphicx}
\usepackage[space]{grffile}
\usepackage{latexsym}
\usepackage{amsfonts,amsmath,amssymb}
\usepackage{url}
\usepackage[utf8]{inputenc}
\usepackage{fancyref}
\usepackage{hyperref}
\hypersetup{colorlinks=false,pdfborder={0 0 0},}
\usepackage{textcomp}
\usepackage{longtable}
\usepackage{multirow,booktabs}
\usepackage{natbib}

\usepackage{bm}
\renewcommand{\vec}[1]{{\bm #1}}

\newcommand{\llnl}{Lawrence Livermore National Laboratory, Livermore, CA 94550, United States}

\newcommand{\harvard}{Department of Chemistry and Chemical Biology, Harvard University, 12 Oxford Street, Cambridge, MA 02138, United States}

\newcommand{\MIT}{Department of Materials Science and
  Engineering, Massachusetts Institute of Technology, Cambridge,
  Massachusetts 02139, United States}

\newcommand{\sanse}{Nano-Bio Spectroscopy Group and European
  Theoretical Spectroscopy Facility (ETSF), Universidad del Pa\'{\i}s
  Vasco CFM CSIC-UPV/EHU-MPC \& DIPC, 20018 San Sebasti\'an, Spain}

\newcommand{\liege}{Unit\'e Nanomat, Universit\'e de Li\`ege, All\'ee du 6 Ao\^ut 17, B-4000 Li\`ege, Belgium}

\newcommand{\sansecomp}{Dept. of Computer Architecture and Technology,
  University of the Basque Country UPV/EHU, M. Lardizabal, 1, 20018
  Donostia-San Sebastian, Spain}

\newcommand{\julich}{Peter-Gr\"unberg Institut and Institute
for Advanced Simulation, Forschungszentrum J\"ulich, D-52425 J\"ulich,
Germany}

\newcommand{\queens}{Atomistic Simulation Centre, School of Mathematics and Physics, Queen's University Belfast, University Road, Belfast BT7 1NN, Northern Ireland, United Kingdom}

\newcommand{\coimbra}{Center for Computational Physics, University of Coimbra, Rua Larga, 3004-516 Coimbra, Portugal}

\newcommand{\bifi}{Institute for Biocomputation and Physics of Complex
  Systems (BIFI) and Zaragoza Center for Advanced Modeling (ZCAM),
  University of Zaragoza, E-50009 Zaragoza, Spain}

\newcommand{\araid}{ARAID Foundation, Mar\'{\i}a de Luna 11, Edificio
  CEEI Arag\'on, Zaragoza E-50018, Spain}

\newcommand{\halle}{Institut f\"ur Physik, Martin-Luther-Universit\"at
  Halle-Wittenberg, Von-Seckendorff-Platz 1, 06120 Halle (Saale),
  Germany}

\newcommand{\mpi}{Max Planck Institute for the Structure and Dynamics of Matter, Hamburg, Germany}

\begin{document}

\title{Real-space grids and the Octopus code as tools for the development of
new simulation approaches for electronic systems}

\author{Xavier Andrade}
\affiliation{\llnl}
\affiliation{\harvard}
 
\author{David A. Strubbe}
\affiliation{\MIT}
 
\author{Umberto De Giovannini}
\affiliation{\sanse}
 
\author{Ask Hjorth Larsen}
\affiliation{\sanse}
 
\author{Micael J. T. Oliveira}
\affiliation{\liege}
 
\author{Joseba Alberdi-Rodriguez}
\affiliation{\sansecomp}
\affiliation{\sanse}
 
\author{Alejandro Varas}
\affiliation{\sanse}
 
\author{Iris Theophilou}
\affiliation{\julich}
 
\author{Nicole Helbig}
\affiliation{\julich}
 
\author{Matthieu Verstraete}
\affiliation{\liege}
 
\author{Lorenzo Stella}
\affiliation{\queens}
 
\author{Fernando Nogueira}
\affiliation{\coimbra}
 
\author{Alan Aspuru-Guzik}
\affiliation{\harvard}
 
\author{Alberto Castro}
\affiliation{\bifi}
\affiliation{\araid}
 
\author{Miguel A. L. Marques}
\affiliation{\halle}
 
\author{Angel Rubio}
\affiliation{\mpi}
\affiliation{\sanse}

% You should use BibTeX and apsrev.bst for references
% Choosing a journal automatically selects the correct APS
% BibTeX style file (bst file), so only uncomment the line
% below if necessary.
\bibliographystyle{apsrev4-1}

\begin{abstract}
Real-space grids are a powerful alternative for the simulation of
electronic systems. One of the main advantages of the approach is the
flexibility and simplicity of working directly in real space where the
different fields are discretized on a grid, combined with competitive
numerical performance and great potential for parallelization. These
properties constitute a great advantage at the time of implementing and
testing new physical models. Based on our experience with the Octopus
code, in this article we discuss how the real-space approach has allowed
for the recent development of new ideas for the simulation of electronic
systems. Among these applications are approaches to calculate response
properties, modeling of photoemission, optimal control of quantum
systems, simulation of plasmonic systems, and the exact solution of the
Schr{\"{o}}dinger equation for low-dimensionality systems.

\end{abstract}

%\pacs{Valid PACS appear here}% PACS, the Physics and Astronomy
                             % Classification Scheme.
%\keywords{Suggested keywords}%Use showkeys class option if keyword
                              %display desired
\maketitle

\section{Introduction} 

The development of theoretical methods for the simulation of electronic systems is an active area of study. This interest has been fueled by the success of theoretical tools like density functional theory (DFT)~\cite{Hohenberg_1964,Kohn_1965} that can predict many properties with good accuracy at a relatively modest computational cost. On the other hand, these same tools are not good enough for many applications~\cite{Cohen_2008}, and more accurate and more efficient methods are required.

Current research in the area covers a broad range of aspects of electronic structure simulations: the development of novel theoretical frameworks, new or improved methods to calculate properties within existing theories, or even more efficient and scalable algorithms. In most cases, this theoretical work requires the development of test implementations to assess the properties and predictive power of the new methods. 

The development of methods for the simulations of electrons requires continual feedback and iteration between theory and results from implementation, so the translation to code of new theory needs to be easy to implement and to modify. This is a factor that is usually not considered when comparing the broad range of methods and codes used by chemists, physicists and material scientists. 

The most popular representations for electronic structure rely on basis sets that usually have a certain physical connection to the system being simulated. In chemistry, the method of choice is to use atomic orbitals as a basis to describe the orbitals of a molecule. When these atomic orbitals are expanded in Gaussian functions, it leads to an efficient method as many integrals can be calculated from analytic formulae~\cite{szabo1996modern}. In condensed-matter physics, the traditional basis is a set of plane waves, which correspond to the eigenstates of a homogeneous electron gas. These physics-inspired basis sets have, however, some limitations. For example, it is not trivial to simulate crystalline systems using atomic orbitals~\cite{Dovesi_2014}, and, on the other hand, in plane-wave approaches finite systems must be approximated as periodic systems using a super-cell approach. 

Several alternatives to atomic-orbital and plane-wave basis sets exist~\cite{White_1989,Tsuchida_1995,Harrison_2004,16008435,Genovese_2011}. One particular approach that does not depend on a basis set uses a grid to directly represent fields in real-space. The method was pioneered by Becke~\cite{Becke_1989}, who used a combination of radial grids centered around each atom. In 1994 Chelikowsky, Troullier and Saad~\cite{Chelikowsky_1994} presented a practical approach for the solution of the Kohn-Sham (KS) equations using uniform grids combined with pseudo-potentials. What made the approach competitive was the use of high-order finite differences to control the error of the Laplacian without requiring very dense meshes. From that moment, several real-space implementations have been presented~\cite{Seitsonen_1995,Hoshi_1995,Gygi_1995,Briggs_1996,Fattebert_1996,Beck_1997,Ono_1999,Beck_2000,Nardelli_2001,Marques_2003,Pask_2005,Kronik_2006,Schmid_2006,Krotscheck_2007,Bernholc_2008,Shimojo_2008,Goto_2009,Enkovaara_2010,Iwata_2010,Sasaki_2011,Ono_2011}.

Discretizing in real-space grids does not benefit from a direct physical connection to the system being simulated. However, the method has other advantages. In first place, a real-space discretization is, in most cases, straight-forward to perform starting from the continuum description of the electronic problem. Operations like integration are directly translated into sums over the grid and differential operators can be discretized using finite differences. In fact, most electronic structure codes must rely on an auxiliary real-space discretization used, for example, for the calculation of the exchange and correlation (xc) term of DFT.

Grids are flexible enough to directly simulate different kinds of systems: finite, and fully or partially periodic. It is also possible to perform simulations with reduced (or increased) dimensionality. Additionally, the discretization error can be systematically and continuously controlled by adjusting the spacing between mesh points, and the physical extension of the grid.

The simple discretization and flexibility of the real space grids makes them an ideal framework to implement, develop and test new ideas. Modern electronic structure codes are quite complex, which means that researchers seldom can write code from scratch, but instead need to resort to existing codes to implement their developments.

From the many codes available, in our experience the real-space code Octopus~\cite{Marques_2003,Castro_2006} provides an ideal framework for theory-development work. To illustrate this point, in this article we will explore some recent advances that have been made in computational electronic structure and that have been developed using the Octopus code as a base. We will pay special attention to the most unusual capabilities of the code, and in particular to the ones that have not been described in previous articles~\cite{Marques_2003,Castro_2006,Andrade_2012}.

\section{The Octopus code}

Octopus was started around 2000 in the group of professor Angel Rubio who, at that moment, was as the University of Valladolid, Spain. The first article using Octopus was published in 2001~\cite{Marques_2001}. Today, the code has grown to 200,000 lines of code. Octopus receives contributions from many developers from several countries and its results have been used for hundreds of scientific publications.

The original purpose of Octopus was to perform real-time time-dependent density functional theory (TDDFT) calculations, a method that had been recently proposed at the time for the calculation of excited-state properties in molecules~\cite{Yabana_1996}. Beyond this original feature, over the time the code has become able to perform many types of calculations of ground-state and excited-state properties. These include most of the standard features of a modern electronic-structure package and some not-so-common capabilities. 

Among the current capabilities of Octopus are an efficient real-time TDDFT implementation for both finite and periodic systems~\cite{Bertsch_2000,Aggarwal_2012}. Some of the research presented in this article is based on that feature, such as the simulation of photoemission, quantum optimal control, and plasmonic systems. The code can also perform molecular-dynamics simulations in the Born-Oppenheimer and Ehrenfest approximations. It also implements a modified Ehrenfest approach for adiabatic molecular dynamics~\cite{Alonso_2008,Andrade_2009} that has favorable scaling for large systems. Octopus can perform linear-response TDDFT calculations in different frameworks; these implementations are discussed in sections~\ref{sec:sternheimer} and~\ref{sec:casida}. For visualization, analysis and post-processing, Octopus can export quantities such as the density, orbitals, the current density, or the time-dependent electron localization function~\cite{Burnus_2005} to different formats, including the required DFT data to perform GW/Bethe-Salpeter calculations with the BerkeleyGW code~\cite{Deslippe_2012}.

Octopus is publicly and freely available under the GPL free/open-source license, this includes all the releases as well as the development version. The code is written using the principles of object oriented programming. This means that the code is quite flexible and modular. It provides a full toolkit for code developers to perform the operations required for the implementation of new approaches for electronic-structure calculations.

In order to control the quality of the package, Octopus uses continuous integration tools. The code includes a set of tests that checks most of the functionality by verifying the calculation results. After a new change is commited to the main repository, a set of servers with different configurations compiles the code and runs a series of short tests. This setup quickly detects most of the problems in a commit, from syntax that a compiler will not accept, to unexpected changes in the results. Every night a more comprehensive set of tests is executed by these same servers. The test-suite framework is quite general and is also successfully in use for the BerkeleyGW \cite{Deslippe_2012} and APE \cite{Oliveira2008} codes.

\section{The Sternheimer formulation of linear-response}
\label{sec:sternheimer}
%David, Xavier

In textbooks, perturbation theory is formulated in terms of sums over
states and response functions. These are useful theoretical
constructions that permit a good description and understanding of the
underlying physics. However, this is not always a good starting point for
numerical applications, since it involves the calculation of a large
number of eigenvectors, infinite sums over these eigenvectors, and functions that depend on two or more spatial variables.

An interesting approach that avoids the problems
mentioned above is the formulation of perturbation theory
in terms of differential equations for the variation of the
wave-functions. In the literature, this is usually called
the Sternheimer equation~\cite{Sternheimer_1951} or density functional perturbation theory (DFPT)~\cite{Baroni_2001}. Although a
perturbative technique, it avoids the use of empty states, and has a
favorable scaling with the number of atoms.

Octopus implements a generalized version of the Sternheimer equation that is
able to cope with both static and dynamic response in and out of
resonance~\cite{Andrade_2007}. The method is suited for linear and
non-linear response; higher-order Sternheimer equations can be
obtained for higher-order variations. For second-order response,
however, we apply the \(2n\,+\,1\) theorem (also known as Wigner's  \(2n\,+\,1\) rule)~\cite{Gonze_1989,Corso_1996} to get the
coefficients directly from first-order response variations.

In the Sternheimer formalism, we consider the response to 
a monochromatic perturbative field \(\lambda
\delta{\hat v}(\vec{r})\cos\left(\omega{t}\right)\). This perturbation induces a variation in the time-dependent KS orbitals, which we denote $\delta\varphi_{n}(\vec r, \omega)$. These variations allow us to calculate response observables, for example, the frequency-dependent polarizability. 

In order to calculate the variations of the orbitals we need to solve a linear equation that only depends on the occupied orbitals (atomic units are used throughout)
%\begin{multline}
\begin{equation}
  \label{eq:sternheimer}
  \left\{\hat H  - \epsilon_n\pm\omega +
    \mathrm{i}\eta\right\}\delta\varphi_{n}(\vec r, \pm\omega) = -\mathrm{\hat P}_c\,\delta{\hat H}(\pm\omega) \varphi_n(\vec r)
  \,,
\end{equation}
where the variation of the time-dependent density, given by
%
%\begin{multline}
\begin{equation}
  \label{eq:varrho}
  \delta{n}(\vec r, \omega) = \sum_k f_k \Big\{
    \left[\varphi_n(\vec{r})\right]^*\delta\varphi_{n}(\vec r, \omega)\\
     + \left[\delta\varphi_{n}(\vec r, -\omega)\right]^*\varphi_n(\vec{r})
  \Big\}\ ,
\end{equation}
%\end{multline}
needs to be calculated self-consistently. The first-order variation of the KS Hamiltonian is
\begin{multline}
  \label{eq:h1}
  \delta{\hat H}(\omega)=
  \delta{\hat v}(\vec{r})
  +\int \mathrm{d}\vec{r}' \frac{\delta{n}(\vec{r}',\omega)}{|\vec{r}-\vec{r}'|}\\
  +\int \mathrm{d}\vec{r}' f_{\rm xc}(\vec r, \vec r', \omega)\,\delta{n}(\vec{r}', \omega)
  \ .
\end{multline}
\(\mathrm{\hat P}_c\) is a projection operator, and
\(\eta\) a positive in\-fi\-ni\-te\-si\-mal, essential to obtain the
correct position of the poles of the causal response function, and
consequently to obtain the imaginary part of the
po\-la\-ri\-za\-bi\-li\-ty and remove the divergences of
  the equation for resonant frequencies. In the usual implementation of DFPT, $\hat P_c = 1 - \sum^{\rm occ}_n \left| \varphi_n \right> \left< \varphi_n \right|$ which effectively removes the components of \(\delta\varphi_{n}(\vec r, \pm\omega)\) in the
subspace of the occupied ground-state wave-functions. In linear
response, these components do not contribute to the variation of the
density.

We have found that it is not strictly necessary to project out the occupied subspace, 
the crucial part is simply to remove the
projection of $\delta \varphi_n$ on $\varphi_n$ (and any other states degenerate with it),
which is not physically meaningful and arises from a phase convention. To fix the phase, it is
sufficient to apply a minimal projector
$\hat P_n = 1 - \sum_m^{\epsilon_m = \epsilon_n} \left| \varphi_m \right> \left< \varphi_m \right|$.
We optionally use this approach to obtain the entire response wavefunction, not just the projection in the
unoccupied subspace, which is needed for obtaining effective masses in $\vec{k} \cdot \vec{p}$ theory. While the full projection can become time-consuming
for large systems, it saves time overall since it increases the condition number of the matrix for the linear solver,
and thus reduces the number of solver iterations required to attain a given precision.

We also have implemented the Sternheimer formalism when non-integer occupations are used, as appropriate for metallic systems. In this case weighted projectors are added to both sides of eq.~(\ref{eq:sternheimer})~\cite{DeGironcoli}. We have generalized the equations to the dynamic case \cite{Strubbethesis}. The modified Sternheimer equation is
\begin{multline}
  \left\{\hat H  - \epsilon_n\pm\omega +
    \mathrm{i}\eta + \sum_m \alpha_m \left| \varphi_m \right> \left< \varphi_m \right| \right\}\delta\varphi_{n}(\vec r, \pm\omega) = \\ 
    -\left[ \tilde{\theta}_{{\rm F},n} - \sum_m \beta_{n,m} \left| \varphi_m \right> \left< \varphi_m \right| \right] \delta{\hat H}(\pm\omega) \varphi_n(\vec r)
  \,,
\end{multline}
where
\begin{align}
\alpha_n &= {\rm max} \left( \epsilon_{\rm F} - 3 \sigma - \epsilon_n, 0 \right)\ , \\
\beta_{n,m} &= \tilde{\theta}_{{\rm F},n} \tilde{\theta}_{n,m} + \tilde{\theta}_{{\rm F},m} \tilde{\theta}_{m,n} + \alpha_m \frac{\tilde{\theta}_{{\rm F},n} - \tilde{\theta}_{n,m}}{\epsilon_n - \epsilon_m \mp \omega} \tilde{\theta}_{m,n}\ ,
\end{align}
$\sigma$ is the broadening energy, and $\tilde{\theta}_{ij}$ is the smearing scheme's approximation to the Heaviside function $\theta \left( \left( \epsilon_i - \epsilon_j \right) / \sigma \right)$.
Apart from semiconducting smearing (\textit{i.e.} the original equation above, which corresponds to the zero temperature limit),
the code offers the standard Fermi-Dirac~\cite{Mermin_1965}, Methfessel-Paxton~\cite{Methfessel_1989}, spline~\cite{Holender_1995}, and cold~\cite{Marzari_1999} smearing schemes.
Additionally, we have developed a scheme for handling arbitrary fractional occupations, which do not have to be defined by a function of the energy eigenvalues \cite{Strubbethesis}.

In order to solve eq.~(\ref{eq:sternheimer}) we use a self-consistent iteration scheme similar to the one used for ground-state DFT. In each iteration we need to solve a sparse linear problem where the operator to invert is the shifted KS Hamiltonian. For real wavefunctions and a real shift (as for the static case), we can use conjugate gradients. When the shift is complex, a non-Hermitian iterative solver is required. We have found that a robust and efficient solver is the quasi-minimal residual (QMR) method~\cite{Freund_1991}.

We can solve for linear response to various different perturbations. The most straight-forward case is the response of a finite system to an electric field \(\mathcal{E}_{i,\omega}\) with frequency \(\omega\) in the direction \(i\), where the perturbation operator is \(\delta \hat v = \hat r_i\)~\cite{Andrade_2007}. In this case the polarizability can be calculated as
\begin{equation}
\alpha_{ij}\left( \omega \right) = - \sum_n^{\rm occ} \left[ \langle \varphi_n \vert \hat{r}_i \vert \frac{\partial \varphi_n}{\partial \mathcal{E}_{j, \omega}} \rangle + \langle \frac{\partial \varphi_n}{\partial \mathcal{E}_{j, -\omega}} \vert \hat{r}_i \vert \varphi_n \rangle \right] \label{eq:sternheimer_polarizability}\ .
\end{equation}
The calculation of the polarizability yields optical response properties (that can be extended to nonlinear response)~\cite{Andrade_2007,Vila_2010} and, for imaginary frequencies, van der Waals coefficients~\cite{Botti_2008}.

It is also possible to use the formalism to compute vibrational properties for finite and periodic systems~\cite{Baroni_2001,Kadantsev_2005}. Currently Octopus implements the calculations of vibrations for finite systems. In this case the perturbation operator is an infinitesimal ionic displacement
$\partial \hat{H}/\partial R_{i \alpha} = \partial \hat{v}_{\alpha}/\partial R_{i \alpha}$, for each direction $i$ and atom $\alpha$.
The quantity to calculate is the dynamical matrix, or Hessian, given by
\begin{multline}
  D_{i \alpha, j \beta} = \frac{\partial^2 E}{\partial R_{i \alpha} \partial R_{j \beta}} = D_{i \alpha, j \beta}^{\rm ion-ion} \\
  - \sum_n^{\rm occ} \left[ \left< \varphi_n \left| \frac{\partial
          \hat{v}_{\alpha}}{\partial R_{i \alpha}} \right|
      \frac{\partial \varphi_n}{\partial R_{j \beta}} \right> + {\rm
      c.c.} \right.\\
      \left. + \delta_{\alpha \beta} \left< \varphi_n \left| \frac{\partial^2 \hat{v}_{\alpha}}{\partial R_{i \alpha} \partial R_{\alpha j}} \right| \varphi_n \right> \right] \label{eq:dynmatrix}
\end{multline}
The contribution from the ion-ion interaction energy is
\begin{align}
    D_{i \alpha, j \beta}^{\rm ion-ion}\! = \!
    \begin{cases}
    \!Z_\alpha \!\sum_{\gamma \ne \alpha}\! Z_\gamma \!\left[
      \frac{\delta_{ij}}{\left|R_\alpha - R_\gamma \right|^3} - 3
      \frac{\left( R_{i \alpha} - R_{i \gamma} \right)}{\left|R_\alpha
          - R_\gamma \right|^4} \right] & \!\!\!\!\alpha = \beta \\
    \!-Z_\alpha Z_\beta \!\left[ \frac{\delta_{ij}}{\left|R_\alpha - R_\beta \right|^3} - 3 \frac{\left( R_{i \alpha} - R_{i \beta} \right)}{\left|R_\alpha - R_\beta \right|^4} \right]              & \!\!\!\!\alpha \ne \beta
    \end{cases}
\end{align}
where $Z_\alpha$ is the ionic charge of atom $\alpha$.
We have found that an alternative expression for the perturbation operator yields more accurate results when discretized. This is discussed in section~\ref{sec:forces}.

Vibrational frequencies $\omega$ are obtained by solving the eigenvalue equation
\begin{equation}
\frac{1}{\sqrt{m_\alpha m_\beta}} D_{i \alpha, j \beta} x_{j \beta} = -\omega^2 x_{j \beta}\ ,
\end{equation}
where \(m_\alpha\) is the mass for ion \(\alpha\). For a finite system of $N$ atoms, there should be 3 zero-frequency translational modes and 3
zero-frequency rotational modes. However, they may appear at positive or imaginary frequencies,
due to the finite size of the simulation domain, the discretization of the grid, and finite precision
in solution of the ground state and Sternheimer equation. Improving convergence brings them closer to zero.

The Born effective charges can be computed from the response of the dipole moment to ionic displacement:
\begin{multline} \label{eq:Born}
  Z^{*}_{i j \alpha} = -\frac{\partial^2 E}{\partial \mathcal{E}_i \partial R_{j \alpha}}
  = \frac{\partial \mu_i}{\partial R_{j \alpha}} \\ 
  = Z_{\alpha} \delta_{ij} - \sum_n^{\rm occ} \left< \varphi_n \left| \hat{r}_i \right| \frac{\partial \varphi_n}{\partial R_{j \alpha}} \right>\,.
\end{multline}
The intensities for each mode for absorption of radiation polarized in direction $i$, which can be used to predict infrared spectra, are calculated by multiplying by the normal-mode eigenvector $x$
\begin{align}
I_i = \sum_{j \alpha} Z^{*}_{ij \alpha} x_{j \alpha}\ .
\end{align}

The Born charges must obey the acoustic sum rule, from translational invariance
\begin{align}
\sum_{\alpha} Z^{*}_{i j \alpha} = Q_{\rm tot} \delta_{ij}\ .
\end{align}
For each $ij$, we enforce this sum rule by distributing the discrepancy equally among the atoms, and thus obtaining corrected Born charges:
\begin{equation}
\tilde{Z}^{*}_{i j \alpha} = Z^{*}_{i j \alpha} + \frac1N\left( Q_{\rm tot} \delta_{ij} - \sum_{\alpha} Z^{*}_{i j \alpha} \right)\ .
\end{equation}
The discrepancy arises from the same causes as the non-zero translational and rotational modes.

The Sternheimer equation can be used in conjunction with $\vec{k} \cdot \vec{p}$ perturbation theory~\cite{Cardona_1966} to obtain
band velocities and effective masses, as well as to apply electric fields via the quantum theory of polarization. In this case the perturbation is a displacement in the \(k\)-point. Using the effective Hamiltonian for the \(k\)-point
\begin{equation}
\hat{H}_{\vec{k}} = \mathrm{e}^{-i \vec{k} \cdot \vec{r}} \hat{H} \mathrm{e}^{i \vec{k} \cdot \vec{r}}\ ,
\end{equation}
the perturbation is represented by the operator
\begin{equation}
  \frac{\partial \hat{H}_{\vec{k}}}{\partial \vec{k}} = -i \nabla + \vec{k} + \left[\hat{v}_{\alpha}, \vec{r} \right]\ ,
\end{equation}
including the effect on the non-local pseudopotentials.
The first-order term gives the band group velocities in a periodic system,
\begin{equation}
  \mathrm{v}_{n\vec{k}} = \frac{\partial \epsilon_{n\vec{k}}}{\partial \vec{n}} 
  = \left< \varphi_{n\vec{k}} \left| \frac{\partial \hat H_{\vec{k}}}{\partial \vec{k}} \right| \varphi_{n\vec{k}} \right>\ .
\end{equation}
%adding in occupied contributions

Inverse effective mass tensors can be calculated by solving the Sternheimer equation for the \(\vec{k}\cdot\vec{p}\) perturbation. The equation is not solved self-consistently, since the variation of \(k\)-point is not a physical perturbation to the system; a converged \(k\)-grid should give the same density even if displaced slightly. The perturbation $\partial \hat{H}_{\vec{k}}/\partial \vec{k}$ is purely anti-Hermitian. We use instead $-i \partial \hat{H}_{\vec{k}}/\partial \vec{k}$ to obtain a Hermitian perturbation, which allows the response to real wavefunctions to remain real.
The effective mass tensors are calculated as follows:
\begin{multline}
  m^{-1}_{ijn\vec{k}} = \frac{\partial^2 \epsilon_{n\vec{k}}}{\partial k_i \partial k_j}
  = \delta_{ij} + \left< \varphi_{n\vec{k}} \left| \frac{\partial \hat{H}_{\vec{k}}}{\partial k_i} \right| \frac{\partial \varphi_{n\vec{k}}}{\partial k_j} \right> + \mathrm{c.c.}
  \\
  + \left< \varphi_{n\vec{k}} \left| \left[ \hat{r}_i, \left[ \hat{r}_j, \hat{v}_{\alpha} \right] \right] \right| \varphi_{n\vec{k}} \right>\ .
\end{multline}

The $\vec{k} \cdot \vec{p}$ wavefunctions can be used to compute the response to electric fields in periodic systems.
In finite systems, a homogeneous electric field can be represented simply via the position operator $\vec{r}$.
However, this operator is not well defined in a periodic system and cannot be used. According to the quantum theory of polarization,
the solution is to replace $\vec{r} \varphi$ with $-i \partial \varphi/\partial k$~\cite{Resta_2007}, and then use this as the perturbation on the right hand side in the Sternheimer equation \cite{Gonze1997}.
While this is typically done with a finite difference with respect to $\vec{k}$ \cite{Corso_1996,QE-2009}, we use an analytic derivative from a previous $\vec{k} \cdot \vec{p}$ Sternheimer calculation.
Using the results in eq.~(\ref{eq:sternheimer_polarizability}) gives a formula for the polarization of the crystal:
\begin{align}
  \alpha_{ij} \left( \omega \right) = i \sum_n^{\rm occ} \left[ \left. \left< \frac{\partial \varphi_{n\vec{k}}}{\partial k_i} \right| \frac{\partial \varphi_{n\vec{k}}}{\partial \mathcal{E}_{j, \omega}} \right> + \left. \left< \frac{\partial \varphi_{n\vec{k}}}{\partial \mathcal{E}_{j, -\omega}} \right| \frac{\partial \varphi_{n\vec{k}}}{\partial k_j} \right> \right]\ .
\end{align}
The polarizability is most usefully represented in a periodic system via the dielectric constant
\begin{align}
  \epsilon_{ij} = \delta_{ij} + \frac{4 \pi}{V} \alpha_{ij}\ ,
\end{align}
where $V$ is the volume of the unit cell. This scheme can also be extended to non-linear response.

We can compute the Born charges from the electric-field response in either finite or periodic systems (as a complementary approach to using the vibrational response):
\begin{multline}
  Z^{*}_{i j \alpha} = -\frac{\partial^2 E}{\partial \mathcal{E}_i \partial R_{j \alpha}} = \frac{\partial F_{j \alpha}}{\partial \mathcal{E}_i} \\
  = Z_\alpha \delta_{ij} - \sum_n^{\rm occ} \left[ \left< \varphi_n \left| \frac{\partial \hat{v}_{\alpha}}{\partial R_{i \alpha}} \right| \frac{\partial \varphi_n}{\partial \mathcal{E}_j} \right> + {\rm c.c.} \right]
\end{multline}
This expression can be evaluated with the same approach as for the dynamical matrix elements, and is easily generalized to non-zero frequency too.
We can also make the previous expression eq. \ref{eq:Born} for Born charges from the vibrational perturbation usable in a periodic system with the replacement $\vec{r} \varphi \rightarrow -i \partial \varphi/\partial k$.

Unfortunately, the $\vec{k} \cdot \vec{p}$ perturbation is not usable to calculate the polarization \cite{Resta_2007}, and a sum over strings of \(k\)-points on a finer grid is required. We have implemented the special case of a $\Gamma$-point calculation for a large super-cell, where the single-point Berry phase can be used \cite{Yaschenko1998}. For cell sizes $L_i$ in each direction, the dipole moment is derived from the determinant of a matrix whose basis is the occupied KS orbitals:
\begin{align}
\mu_i = - \frac{L_i}{2 \pi} \mathrm{Im}\ {\rm ln}\ {\rm det}\ \left< \varphi_n \left| e^{- 2 \pi i x_i/L_i} \right| \varphi_m \right>\ .
\end{align}

% magnetic is non-self-consistent if wfns are real.

\section{Magnetic response and gauge invariance in real-space grids}
%Miguel, Xavier

In the presence of a magnetic field \(\vec{B}(\vec{r}, t)\), generated
by a vector potential \(\vec{A}(\vec{r}, t)\), additional terms describing the coupling of 
the electrons to the magnetic field must be included in the Hamiltonian
\begin{equation}
  \label{eq:magnetic}
  \hat{H} = \frac12\left({\hat{\vec{p}}} -\frac1c{\vec{A}}\right)^2 + \hat{v} + \vec{B}\cdot{\hat{\vec{S}}}\ .
\end{equation}
The first part describes the orbital interaction with the field, and the
second one is the Zeeman term that represents the coupling of the
electronic spin with the magnetic field.

As our main interest is the evaluation of the magnetic susceptibility, 
in the following, we consider a perturbative uniform static magnetic
field \(\vec{B}\) applied to a finite system with zero total spin. In the Coulomb gauge the
corresponding vector potential, \(\vec{A}\), is given as
\begin{equation}
  \label{eq:unifmag}
  \vec{{A}}(\vec{r}) = \frac12\vec{B}\times{\vec{r}}\ .
\end{equation}
In orders of \(\vec{B}\) the perturbing potentials are
\begin{equation}
  \label{eq:mag1}
  \delta\hat{v}_i^{\mathrm{mag}} = \frac1{2c}({\vec{r}}\times{\hat{\vec{p}}})_i = \frac1{2c}\hat{L}_i\ ,
\end{equation}
with \(\vec{\hat{L}}\) the angular momentum operator, and
\begin{equation}
  \label{eq:mag2}
  \delta^2\hat{v}_{ij}^{\mathrm{mag}} = \frac1{8c^2}(\delta_{ij}r^2 - r_i r_j)\ .
\end{equation}

The induced magnetic moment can be expanded in terms of the external magnetic field which, to first order, reads
\begin{equation}
  \label{eq:bchi}
  m_i=m^0_i+\sum_j\chi_{ij}B^{ext}_j\ ,
\end{equation}
where \(\vec{\chi}\) is the magnetic susceptibility tensor. For finite systems the
permanent magnetic moment can be calculated directly from the
ground-state wave-functions as
\begin{equation}
  \label{eq:magneticmoment}
  \vec{m}^0=
  \sum_n\langle\varphi_n|\delta{\hat{\vec{v}}}^{\mathrm{mag}}|\varphi_{n}\rangle\ .
\end{equation}
For the susceptibility, we need to calculate the first-order response
functions in the presence of a magnetic field. This can be done in
practice by using the magnetic perturbation, eq.~(\ref{eq:mag1}), in
the Sternheimer formalism described in 
section \ref{sec:sternheimer}. If the system is time-reversal symmetric, since the perturbation is anti-symmetric under time-reversal (anti-Hermitian), it does not induce a change
in the density and the Sternheimer equation does not need to be solved
self-consistently. From there we find
\begin{multline}
  \label{eq:susc}
  \chi_{ij} = \sum_n\Big[
    \langle{\varphi_n}|\delta\hat{v}^{\mathrm{mag}}_j|{\delta\varphi_{n,i}}\rangle + {\rm c.c.}
    %+\langle{\delta\varphi_{n\,,\,i}}|\delta\hat{v}^{\mathrm{mag}}_j|{\varphi_{n}}\rangle\\
    +\langle{\varphi_n}|\delta^2\hat{v}^{\mathrm{mag}}_{ij}|{\varphi_{n}}\rangle
  \Big]\ .
\end{multline}
Before applying this formalism in a calculation, however, we must make sure that our calculation is gauge invariant.

In numerical implementations, the gauge freedom in choosing the vector
potential might lead to poor convergence with the quality of the
discretization, and to a dependence of the magnetic response on the
origin of the simulation cell. In other words, an arbitrary
translation of the molecule could introduce an nonphysical change in
the calculated observables. This broken gauge-invariance is well known
in molecular calculations with all-electron methods that make use of
localized basis sets. In this case,
the error can be traced to the finite-basis-set representation of the
wave-functions~\cite{Wolinski_1990,Schindler_1982}. A simple measure of the error
is to check for the fulfillment of the hyper-virial
relation~\cite{Bouman_1977}.
\begin{equation}
  \label{eq:hypervirial}
  i\langle\varphi_j\vert{\hat{\vec{p}}}\vert\varphi_n\rangle = 
  (\epsilon_n-\epsilon_j)\langle\varphi_j\vert{\hat{\vec{r}}}\vert\varphi_n\rangle\ ,
\end{equation}
where \(\epsilon_n\) is the eigenvalue of the state \(\varphi_n\).

When working with a real-space mesh, this problem also appears, though
it is milder, because the standard operator representation in the grid is
not gauge-invariant. In this case the error can be controlled by
reducing the spacing of the mesh. On the other hand, real-space grids usually require the use of the pseudo-potential approximation, where the electron-ion interaction is
described by a non-local potential
\(\hat v_{\text{nl}}\). This, or any other non-local
potential, introduces a fundamental problem when describing the
interaction with magnetic fields or vector potentials in general. To
preserve gauge invariance, this term must be adequately coupled to the
external electromagnetic field, otherwise the results will strongly
depend on the origin of the gauge. For example, an extra term has to be
included in the hyper-virial expression, eq.~(\ref{eq:hypervirial}), resulting in
\begin{equation}
 \label{eq:hypervirial_nl}
 i\langle\varphi_j\vert\hat{\vec{p}}\vert\varphi_n\rangle = 
 (\epsilon_n-\epsilon_j)\langle\varphi_j\vert\hat{\vec{r}}\vert\varphi_n\rangle +
 \langle\varphi_j\vert[\hat{\vec{r}},\hat{v}_{\text{nl}}]\vert\varphi_n\rangle\ .
\end{equation}

In general, the gauge-invariant non-local potential is given by
\begin{equation}
 \label{eq:gauge-methods}
 \langle \vec{r} | \hat{v}^{\vec{{A}}}_{\text{nl}} | \vec{r}' \rangle =
 \langle \vec{r} | \hat{v}_{\text{nl}}| \vec{r}' \rangle \mathrm{exp}\left(\frac{i}{c}\int_{\vec{r}}^{\vec{r}'} \vec{A}(\vec{x},t) \cdot \mathrm{d}\vec{x}\right)
 \ .
\end{equation}
The integration path can be any one that connects the two
points \(\vec{r}\) and \(\vec{r}'\), so an infinite number of choices
is possible. 

In order to calculate the corrections required to the magnetic perturbation operators, we use two different integration paths that have been
suggested in the literature. The first was proposed by Ismail-Beigi,
Chang, and Louie (ICL)~\cite{Ismail_Beigi_2001} who give the
following correction to the first-order magnetic perturbation term
\begin{equation}
  \label{eq:iclmag1}
  \delta \hat{\vec{v}}^{\mathrm{ICL}} = \delta \vec{\hat{v}}^{\mathrm{mag}} - \frac{i}{2c}\vec{\hat{r}}\times\left[\hat{\vec{r}}, v_{\text{nl}}\right]\ ,
\end{equation}
and a similar term for the second-order perturbation. Using a different integration path, Pickard and Mauri~\cite{Pickard_2003} proposed the GIPAW method, that has the form
%GIPAW
%
\begin{equation}
  \label{eq:gipawmag1}
  \delta \hat{\vec{v}}^{\mathrm{GIPAW}} = \delta \hat{\vec{v}}^{\mathrm{mag}} - \frac{i}{2c}\sum_\alpha\vec{R}_\alpha\times\left[\hat{\vec{r}}, \hat{v}^\alpha_{\text{nl}}\right]\ ,
\end{equation}
where \(\vec{R}_\alpha\) and \(\hat{v}^\alpha_{\text{nl}}\) are, respectively, the position and non-local potential of atom \(\alpha\).
With the inclusion of either one of these methods, both implemented in Octopus, we recover gauge invariance in our formalism when pseudo-potentials are used. This allows us to predict the magnetic susceptibility and other properties that depend on magnetic observables, like optical activity~\cite{Varsano_2009}.

\begin{figure}[h!]
\begin{center}
\includegraphics*[width=0.9\columnwidth]{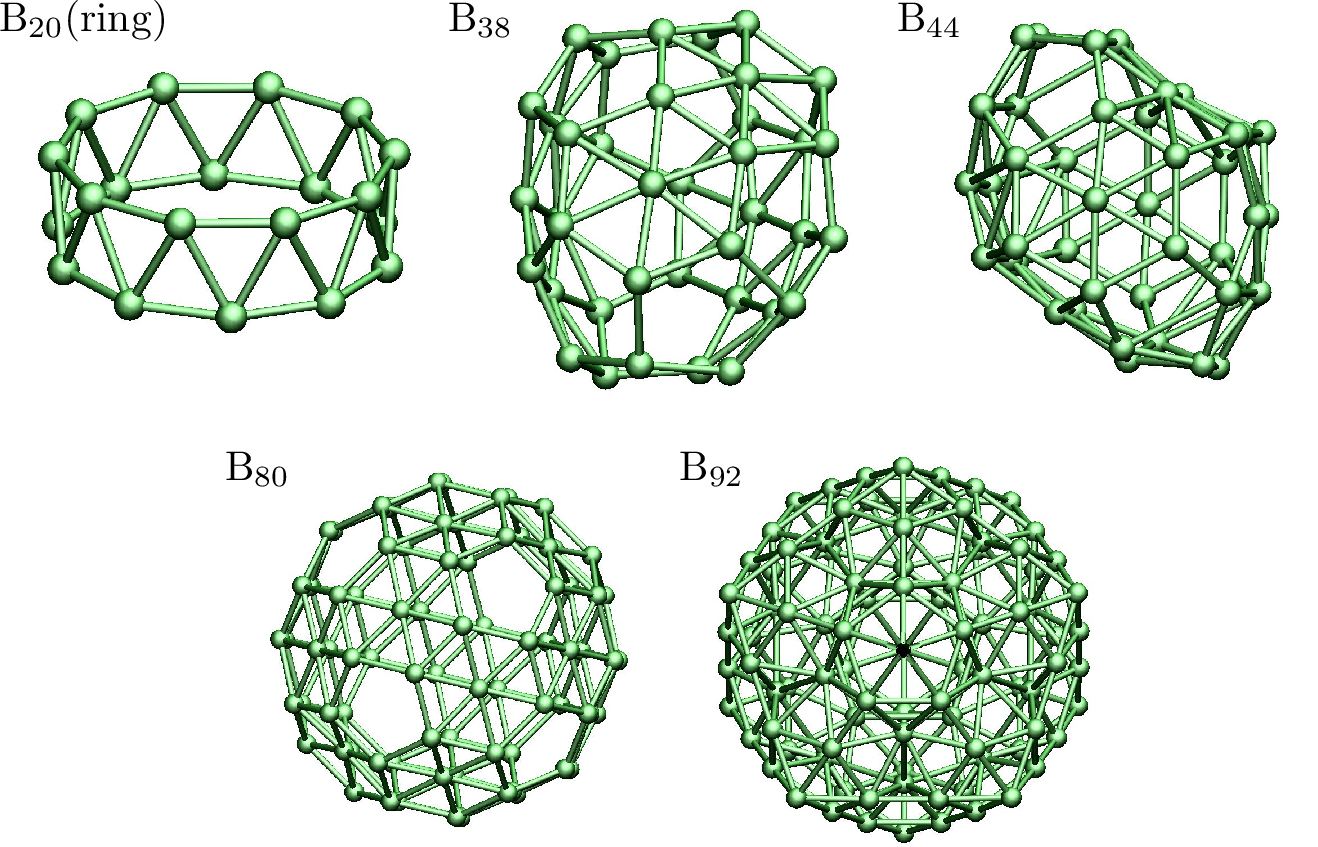}
\caption{\label{fig:boronfullerenes} Structures of boron cages whose magnetic susceptibilities are given in table~\ref{tab:boronmagnetic}.}
\end{center}
\end{figure}

A class of systems with interesting magnetic susceptibilities are fullerenes. For example, it is known that the C\(_{60}\) fullerene has a very small magnetic susceptibility due to the cancellation of the paramagnetic and diamagnetic responses~\cite{Haddon_1991,Haddon_1995}. Botti~\emph{et al.}~\cite{Botti_2009} used the real-space implementation of Octopus to study the magnetic response of the boron fullerenes depicted in Fig.~\ref{fig:boronfullerenes}. As shown in table~\ref{tab:boronmagnetic}, they found that, while most clusters are diamagnetic, B\(_{80}\) is paramagnetic, with a strong cancellation of the paramagnetic and diamagnetic terms.

\begin{table}
\centering
\begin{tabular}{lc}
\mbox{} & $\bar\chi$ \\ \hline
B$_{20}$     & -250.2  \\
B$_{38}$     & -468.3  \\
B$_{44}$     & -614.4  \\
B$_{80}$     &  219.3  \\
B$_{92}$     & -831.3 
\end{tabular}
\caption{
\label{tab:boronmagnetic}
Calculated magnetic susceptibilities ($\chi$
in cgs\,ppm/mol) per number of boron atoms
for the selected boron clusters shown in Fig.~\ref{fig:boronfullerenes}. Results from Ref.~\cite{Botti_2009}}
\end{table}

\section{Linear response in the electron-hole basis}
\label{sec:casida}

An alternate approach to linear response is not to solve for the response function but rather for its poles (the excitation energies $\omega_k$) and residues (\textit{e.g.} electric dipole matrix elements $\vec{d}_k$) \cite{Strubbe2012}. The polarizability is given by
\begin{align}
  \alpha_{ij} \left( \omega \right) = \sum_k \left[ \frac{\left( \hat{i} \cdot \vec{d}_k \right)^{*} \left( \hat{j} \cdot \vec{d}_k \right)}{\omega_k - \omega - i \delta}
    + \frac{\left( \hat{i} \cdot \vec{d}_k \right)^{*} \left( \hat{j} \cdot \vec{d}_k \right)}{\omega_k + \omega + i \delta} \right]
\end{align}
and the absorption cross-section is
\begin{align}
\sigma_{ij} \left( \omega \right) = \frac{4 \pi \omega}{c} \tilde{\alpha} \  \mathrm{Im}\ \alpha_{ij} \left( \omega \right)\ ,
\end{align}
where $\tilde{\alpha}$ is the fine-structure constant.
%and the oscillator strengths are
%\begin{align}
%f_k = \frac{2 m \omega_k}{\hbar^2} \left| d_k \right|^2
%\end{align}
The simplest approximation to use is the random-phase approximation (RPA), in which the excitation energies are given by the differences of unoccupied and occupied KS eigenvalues, $\omega_{c v} = \epsilon_c - \epsilon_v$. The corresponding dipole matrix elements are $\vec{d}_{cv} = \left< \varphi_c \left| \vec{r} \right| \varphi_v \right>$ \cite{Onida2002}. (As implemented in the code, this section will refer only to the case of a system without partially occupied levels.)

The RPA is not a very satisfactory approximation, however. The full solution within TDDFT is given by a non-Hermitian matrix eigenvalue equation, with a basis consisting of both occupied-unoccupied ($v \rightarrow c$) and unoccupied-occupied ($c \rightarrow v$) KS transitions. The equation reads as
\begin{align}
  \left[ \begin{array}{c|c}
               A      & B \\ \hline
               -B^{*} & -A
    \end{array} \right] \vec{x} = \omega \vec{x}\ ,
\end{align}
where the $A$ matrices couple $v \rightarrow c$ transitions among themselves and $c \rightarrow v$ among themselves, while the $B$ matrices couple the two types of transitions. They have the form~\cite{Onida2002}
\begin{align}
  \left< \varphi_{c'} \right| \left< \varphi_{v'} \right| A \left| \varphi_c \right> \left| \varphi_v \right> &=
  \left( \epsilon_c - \epsilon_v \right) \delta_{cc'} \delta_{vv'} \\ \nonumber
  &+ \left< \varphi_{c'} \right| \left< \varphi_{v'} \right| \hat{v}_{\rm c} + \hat{f}_{\rm xc} \left| \varphi_c \right> \left| \varphi_v \right>\ , \\
  \left< \varphi_{c'} \right| \left< \varphi_{v'} \right| B \left| \varphi_c \right> \left| \varphi_v \right> &=
  \left< \varphi_{c'} \right| \left< \varphi_{v'} \right| \hat{v}_{\rm c} + \hat{f}_{\rm xc} \left| \varphi_c \right> \left| \varphi_v \right>\ .
\end{align}
where $\hat{v}_{\rm c}$ is the Coulomb kernel, and $\hat{f}_{\rm xc}$ is the exchange-correlation kernel (currently only supported for LDA-type functionals in Octopus).

We do not solve the full equation in Octopus, but provide a hierarchy of approximations. An example calculation for the N$_2$ molecule with each theory level is shown in Table \ref{tab:nitrogen_casida}.
The lowest approximation we use is RPA.  The next is the single-pole approximation of Petersilka \textit{et al.} \cite{Petersilka1996},
in which only the diagonal elements of the matrix are considered.
Like in the RPA case, the eigenvectors and dipole matrix elements are simply the KS transitions. The positive eigenvalues are $\omega_{cv} = \epsilon_c - \epsilon_v + A_{cvcv}$.
This can be a reasonable approximation when there is little mixing between KS transitions,
but generally fails when there are degenerate or nearly degenerate transitions.

A next level of approximation is the Tamm-Dancoff approximation to TDDFT \cite{Hirata1999}
in which the B blocks are neglected and thus we need only consider the
occupied-unoccupied transitions. The matrix equation is reduced to a Hermitian problem
of half the size of the full problem:
\begin{align}
A \vec{x} = \omega \vec{x}\ .
\end{align}
Interestingly, the Tamm-Dancoff approximation is often found to give superior results to
the full solution, for example for molecular potential-energy surfaces
%cite one of the forces papers? e.g. Tsukagoshi, Sitt, Hutter etc.
or when hybrid functionals are used, which can suffer from a ``triplet instability''
in which the lowest triplet state is lower in energy than the ground state \cite{Casida2009}.
The dipole matrix elements are now a superposition of the KS ones:
\begin{align}
\vec{d}_k = \sum_{cv} \vec{d}_{cv}\, x_{cv}\ .
\end{align}

When the wavefunctions are real, the full problem can be collapsed into a Hermitian one of the same
size as the Tamm-Dancoff matrix, known as Casida's equation \cite{Casida1995,Jamorski1996}.
\begin{align}
  \left( \epsilon_c - \epsilon_v \right)^2 \delta_{cc'} \delta_{vv'} + 2 \sqrt{\epsilon_{c'} - \epsilon_{v'}}
  B_{cvc'v'} \sqrt{\epsilon_c - \epsilon_v} = \omega^2 x_{cv}\,.
\end{align}
The dipole matrix elements are
\begin{align}
\vec{d}_k = \sum_{cv} \vec{d}_{cv}\, x_{cv}\, \sqrt{\frac{\epsilon_c - \epsilon_v}{\omega_k}}\ .
\end{align}

An alternate approach for finding excitation energies is to look for many-body eigenstates
of the DFT Hamiltonian which are orthogonal to the ground state. In the ``second-order constrained
variational'' or CV(2) theory \cite{Ziegler2009},
second-order perturbation theory from the ground-state density
yields equations quite similar to the linear-response approach, despite their different origin:
\begin{align}
  \left[ \begin{array}{c|c}
               A      & B \\ \hline
               B^{*} & -A
    \end{array} \right] \vec{x} = \omega \vec{x}\ .
\end{align}
We implement the case of real wavefunctions and eigenvectors, in which case (as for Casida's equation) a Hermitian matrix equation
for only the occupied-unoccupied transitions can be written:
\begin{align}
\left( A + B \right) \vec{x} = \omega \vec{x}\ .
\end{align}
The Tamm-Dancoff approximation to these equations is identical to the ordinary TDDFT Tamm-Dancoff approximation.

Note that all the levels of theory we have discussed use the same Coulomb and $\hat{f}_{\rm xc}$ matrix elements, so the code can calculate the results for multiple levels of theory with a small extra effort. We can also consider alternative perturbations in this framework beyond the dipole approximation for properties such as inelastic X-ray scattering \cite{Sakko2010}.

\begin{table}
\centering
\begin{tabular}{ccccc|c}
RPA & Petersilka & TDA & Casida & CV(2) & Exp't \\ \hline
8.234    & 9.421      &  9.343       & 9.254  & 9.671       & 9.309 \\
8.234    & 9.421      &  9.343       & 9.254  & 10.279      & 9.309 \\
9.671    & 9.671      &  9.671       & 9.671  & 10.279      & 9.921 \\
9.671    & 10.241     &  10.237      & 10.221 & 10.792      & 10.270 \\
9.671    & 10.245     &  10.241      & 10.224 & 10.801      & 10.270 \\
9.671    & 11.028     &  10.931      & 10.921 & 11.077      & 12.199 \\
\end{tabular}
\caption{
\label{tab:nitrogen_casida}
The first 6 excitation energies (in eV) for the N$_2$ molecule with different approximations to TDDFT in the electron-hole basis: the random phase approximation (RPA), Petersilka, Tamm-Dancoff approximation (TDA), Casida and CV(2). The VWN LDA parametrization~\cite{Vosko_1980} was used for the exchange-correlation functional, the bond length is 1.098 \AA, the real-space grid was a sphere of radius 7.4 \AA~with spacing 0.16 \AA, and 16 unoccupied states were used. The experimental data is from Ref.~\cite{Ben-Shlomo_1990}.}
\end{table}

For a non-spin-polarized system, the excitations separate into a singlet and a triplet subspace,
which are superpositions of singlet and triplet KS transitions:
\begin{align}
  \varphi ^{\rm S} = \frac{1}{\sqrt{2}} \left( \varphi_{c \uparrow} \varphi_{v \uparrow}
  + \varphi_{c \downarrow} \varphi_{v \downarrow} \right)\ , \\
  \varphi ^{\rm T} = \frac{1}{\sqrt{2}} \left( \varphi_{c \uparrow} \varphi_{v \uparrow}
  - \varphi_{c \downarrow} \varphi_{v \downarrow} \right)\ .
\end{align}
The signs are reversed from the situation for a simple pair of electrons, since we are instead dealing
with an electron and a hole. There are of course two other triplet excitations ($m = \pm 1$) which are degenerate with the $m = 0$ one above.
Rather than performing spin-polarized ground-state and linear-response calculations, we can use
the symmetry between the spins in a non-spin-polarized system to derive a form of the kernel to use
in obtaining singlet and triplet excitations~\cite{Onida2002}
\begin{align}
  \left< \varphi^{\rm S} \right| \hat{v}_{\rm c} + \hat{f}_{\rm xc}
  \left| \varphi^{\rm S} \right> 
  & = \left< \varphi \right| \hat{v}_{\rm c} + \hat{f}_{\rm
    xc}^{\uparrow \uparrow} + \hat{f}_{\rm xc}^{\uparrow \downarrow}
    \left| \varphi \right> \nonumber \\
  & = \left< \varphi \right| \hat{v}_{\rm c} + 2 \hat{f}_{\rm xc}
  \left| \varphi \right> \\
  \left< \varphi^{\rm T} \right| \hat{v}_{\rm c} + \hat{f}_{\rm xc}
  \left| \varphi^{\rm T} \right>
  &= \left< \varphi \right| \hat{f}_{\rm xc}^{\uparrow \uparrow} - \hat{f}_{\rm xc}^{\uparrow \downarrow} \left| \varphi \right>\ .
\end{align}
These kernels can be used in any of the levels of theory above: RPA, Petersilka, Tamm-Dancoff, Casida, and CV(2).
The corresponding electric dipole matrix elements are as in the spin-polarized case for singlet excitations.
For triplet excitations, they are identically zero, and only higher-order electromagnetic processes can excite them.

There are three main steps in the calculation: calculation of the matrix, diagonalization of the matrix, and calculation
of the dipole matrix elements. The first step generally takes almost all the computation time, and is the most important
to optimize. Within that step, the Coulomb part (since it is non-local) is much more time-consuming than the $\hat{f}_{\rm xc}$
part. We calculate it by solving the Poisson equation (as for the Hartree potential) for each column of the matrix,
to obtain a potential P for the density $\varphi_c \left( r \right)^{*} \varphi_v \left( r \right)$, and then for each row
computing the matrix element as
\begin{align}
  \left< \varphi_{c'} \varphi_{v'} \right| v \left| \varphi_c \varphi_v \right>
  = \int \mathrm{d}\vec{r}\, \varphi_{c'}(\vec{r}) \varphi_{v'}( \vec{r}) P \left[ \varphi_c \varphi_v \right]\ .
\end{align}

\begin{figure}[h!]
\begin{center}
\includegraphics[width=0.9\columnwidth]{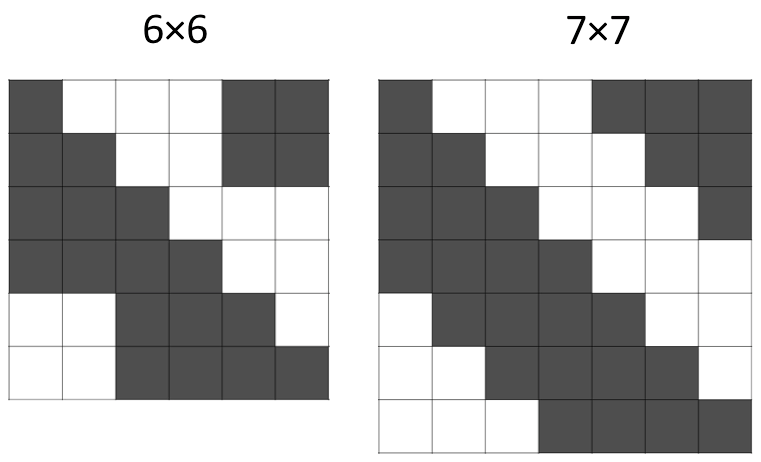}
\caption{\label{fig:casida_parallelization}
Distribution of matrix elements to be calculated among the columns, using Hermiticity of the response matrix. The columns are then distributed among the available MPI groups for electron-hole parallelization. The number of matrix elements to be calculated per column is equal for an odd size, and uneven for an even size. }
\end{center}
\end{figure}

Our basic parallelization strategy for computation of the matrix elements is by domains, as discussed in section~\ref{sec:parallelization},
but we add an additional level of
parallelization here over occupied-unoccupied pairs. We distribute the columns of the matrix, and do not distribute
the rows, to avoid duplication of Poisson solves. We can reduce the number of matrix elements to be computed by
almost half using the Hermitian nature of the matrix, \textit{i.e.} $M_{cv,c'v'} = M_{c'v',cv}^{*}$. If there are $N$
occupied-unoccupied pairs, there are $N$ diagonal matrix elements, and the $N \left( N - 1 \right) / 2$ remaining
off-diagonal matrix elements are distributed as evenly as possible among the columns. If $N-1$ is even,
there are $\left( N - 1 \right) / 2$ for each column; if $N-1$ is odd, half of the columns have $N/2 - 1$ and
half have $N/2$. See Fig. \ref{fig:casida_parallelization} for examples of the distribution. The columns then are assigned to the available
processors in a round-robin fashion. The diagonalization step is performed by direct diagonalization with LAPACK~\cite{LAPACK}
in serial; since it generally accounts for only a small part of the computation time, parallelization of this step is not very important. The final step is calculation of
the dipole matrix elements, which amounts to only a small part of the computation time, and uses only domain parallelization.
Note that the triplet kernel lacks the Coulomb term, and so is considerably faster to compute.

Using the result of a calculation of excited states by one of these methods, and a previous calculation
of vibrational modes with the Sternheimer equation, we can compute forces in each excited state, which can be used 
for excited-state structural relaxation or molecular dynamics \cite{Strubbe_forces}. Our formulation allows us to do this without introducing any extra summations over empty states, unlike previous force implementations \cite{Sitt2007,Tsukagoshi2012,Hutter2003}.
The energy of a given excited state $k$ is a sum of the ground-state energy and the excitation energy: $E_k = E_0 + \omega_k$.
The force is then given by the ground-state force, minus the derivative of the excitation energy:
\begin{align}
  F^{k}_{i \alpha} = -\frac{\partial E_k}{\partial R_{i \alpha}} = F_{i \alpha} - \frac{\partial \omega_k}{\partial R_{i \alpha}}\ .
\end{align}
Using the Hellman-Feynman Theorem we find the last term without introducing any additional sums over unoccupied states.
In the particular case of the Tamm-Dancoff approximation we have
\begin{equation}
  \frac{\partial \omega_k}{\partial R_{i \alpha}} = \left< x_k \left| \frac{\partial \hat{A}}{\partial R_{i \alpha}} \right| x_k \right>\ ,
  \end{equation}
  and
  \begin{multline}
  \left<\varphi_c \varphi_v \left| \frac{\partial \hat{A}}{\partial R_{i \alpha}} \right| \varphi_{c'} \varphi_{v'} \right> =
  \left<\varphi_c \left| \frac{\partial \hat{H}}{\partial R_{i \alpha}} \right| \varphi_{c'} \right> \delta_{vv'}  \\
  - \left<\varphi_v \left| \frac{\partial \hat{H}}{\partial R_{i \alpha}} \right| \varphi_{v'} \right> \delta_{cc'}
  + \left<\varphi_c \varphi_v \left| \hat{K}_{\rm xc} \frac{\partial \rho}{\partial R_{i \alpha}} \right| \varphi_{c'} \varphi_{v'} \right>\,.
\end{multline}
Analogous equations apply for the difference of eigenvalues, Petersilka, and CV(2) theory levels. (The slightly more complicated Casida case has not yet been implemented.)
The Coulomb term, with no explicit dependence on the atomic positions, does not appear, leading to a significant savings in computational time
compared to the calculation of the excited states.

\section{Forces and geometry optimization on real-space grids}
\label{sec:forces}
%Xavier, Alejandro

A function represented on a real-space grid is not invariant under translations as one would expect from a physical system. The potential of an atom sitting on top of a grid point might be slightly different from the potential of the same atom located between points. This implies that a rigid displacement of the system produces an artificial variation of the energy and other properties. If we plot the energy of the atom as a function of this rigid displacement, the energy shows an oscillation that gives this phenomenon the name of the ``egg-box effect''.

The egg-box effect is particularly problematic for calculations where the atoms are allowed to move, for example to study the dynamics of the atoms (molecular dynamics) or to find the minimum energy configuration (geometry optimization).

In Octopus we have studied several schemes to control the egg-box effect~\cite{Andrade2010thesis}. The first step is to use pseudo-potential filtering to eliminate Fourier components of the potential that cannot be represented on the grid~\cite{Tafipolsky_2006}.

\begin{figure}[h!]
\begin{center}
\includegraphics[width=0.9\columnwidth]{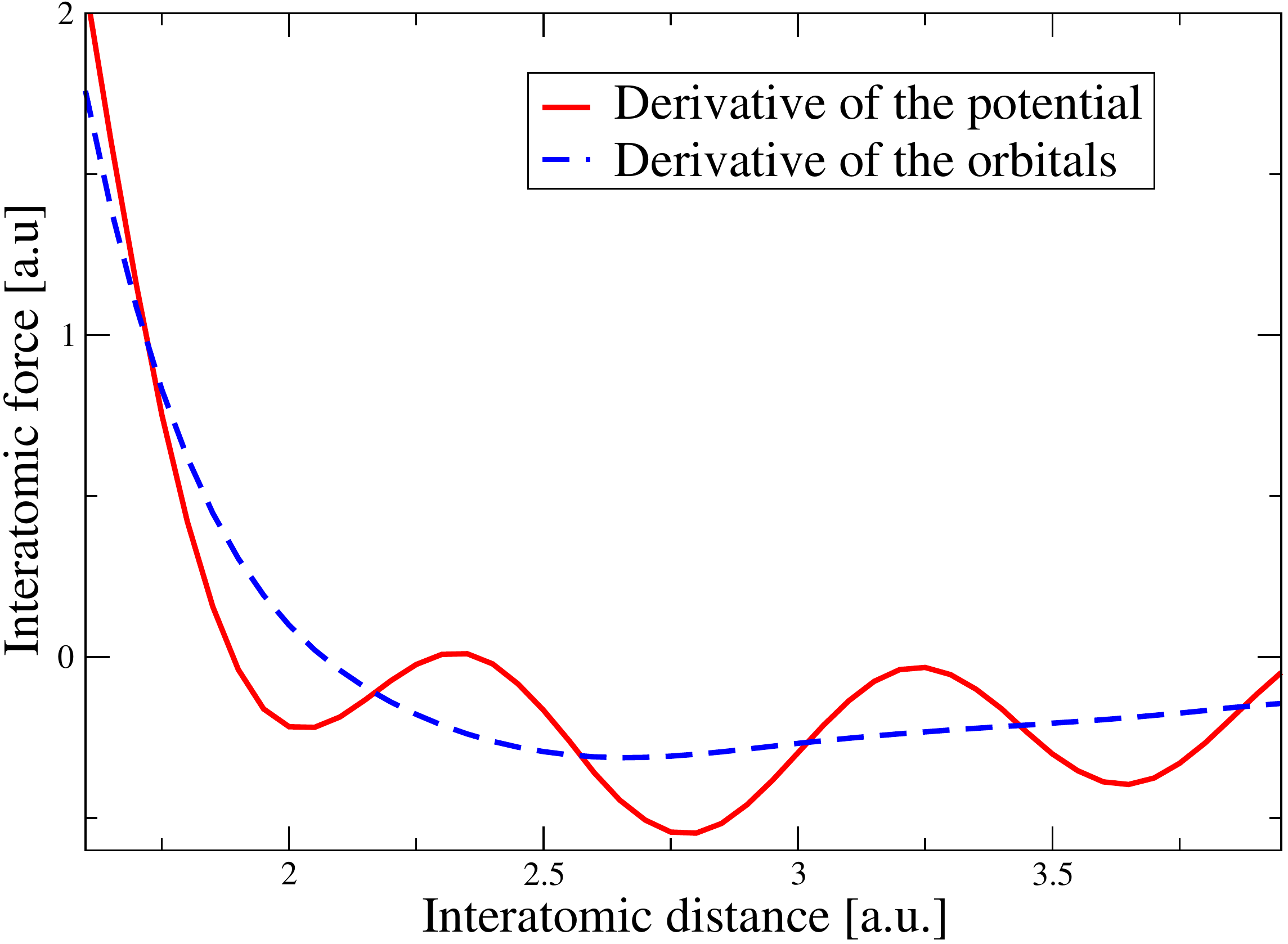}
\caption{\label{fig:n2forces}
Calculation of the interatomic force for N$_2$. Solid (red) line: force calculated from the derivative of the ionic potential with respect to the atomic position. Segmented (blue) line: force calculated from spatial derivatives of the molecular orbitals. Grid spacing of 0.43 Bohr. }
\end{center}
\end{figure}

Additionally, we have found a formulation for the forces that reduces the spurious effect of the grid on the calculations. One term in the  forces is the expectation value of
the derivative of the ionic potential with respect to the ionic position
\(\vec{R_{\alpha}}\), which can be evaluated as
\begin{equation}
  \label{eq:forcespot}
  \vec{F}_{\alpha} = \vec{F}_{\alpha}^{\mathrm{ion-ion}}
  -\sum_{n} \left< \varphi_{n} \left| \frac{\partial \hat{v}_{\alpha}}{\partial \vec{R}_{\alpha}} \right| \varphi_{n} \right>\ .
    %\int \mathrm{d}\vec{r}
    %\varphi^*_{n}(\vec{r})\frac{\partial v_{\alpha}(\vec{r}-
    %\vec{R}_\alpha)}{\partial \vec{R}_{\alpha}}\varphi_{n}(\vec{r})\ .
\end{equation}
(For simplicity, we consider only local potentials here, but the results are valid for non-local potentials as well.)
This term can be rewritten such that it does not include the
derivative of the ionic potential \(v_{\alpha}\), but the gradient of the orbitals with respect to the electronic coordinates~\cite{hirose2005first}:
\begin{equation}
  \label{eq:forcesgrad}
  \vec{F}_\alpha = \vec{F}_\alpha^{\mathrm{ion-ion}}+\sum_{n} \left[ \left< \frac{\partial \varphi_{n}}{\partial r_{\vec{r}}} \left| \hat{v}_{\alpha} \right| \varphi_{n} \right> + \mathrm{c.c.} \right]\,.
    %\int \mathrm{d}\vec{r}
    %\frac{\partial \varphi^*_{n}(\vec{r})}{\partial \vec{r}}v_{\alpha}(\vec{r}-
    %\vec{R}_\alpha)\varphi_{n}(\vec{r}) + \mathrm{c.c.}\,.
\end{equation}
The first advantage of this formulation is that it is easier to implement than eq.~\eqref{eq:forcespot}, as it does not require the derivatives of the potential, which can be quite complex and difficult to code, especially when relativistic corrections are included. 
However, the main benefit of using eq.~\eqref{eq:forcesgrad} is that it is more precise when discretized on a grid, as the orbitals are smoother than the ionic potential. We illustrate this point in Fig.~\ref{fig:n2forces}, where the forces obtained with the two methods
are compared. While taking the derivative of the atomic potential
gives forces with a considerable oscillation due to the grid, using
the derivative of the orbitals gives a force that is considerably smoother.

This alternative formulation of the forces can be extended to obtain the second-order derivatives of the energy with respect to the atomic displacements~\cite{Andrade2010thesis}, which are required to calculate vibrational properties as discussed in section~\ref{sec:sternheimer}. In general, the perturbation operator associated with an ionic displacement can be written as

\begin{equation}
  \label{eq:ionicpertmod}
  \frac{\partial {v}_\alpha(\vec{r}-
    \vec{R}_\alpha)}{\partial R_{i\alpha}} = 
  -{v}_{\alpha}(\vec{r}-
    \vec{R}_\alpha)\frac{\partial}{\partial r_i}
  -\frac{\partial}{\partial
    r_i}{v}_\alpha(\vec{r}-
    \vec{R}_\alpha)\ .
\end{equation}
Using this expression, the terms of the dynamical matrix, eq.~(\ref{eq:dynmatrix}), are evaluated as
\begin{multline}
  \left< \varphi_n \left| \frac{\partial \hat{v}_{\alpha}}{\partial R_{i\alpha}} \right| \frac{\partial \varphi_n}{\partial R_{j \beta}} \right> = -\left[
 \left< \varphi_n \left| \hat{v}_{\alpha} \right| \frac{\partial^2 \varphi_n}{\partial R_{j \beta} \partial r_i} \right> \right.\\ + \left. \left< \frac{\partial \varphi_n}{\partial r_i}\left| \hat{v}_{\alpha} \right| \frac{\partial \varphi_n}{\partial R_{j \beta}} \right>\right] + {\rm c.c.}\ ,
\end{multline}
and
\begin{multline} 
  \left< \varphi_n \left| \frac{\partial^2 \hat{v}_{\alpha}}{\partial R_{i\alpha} \partial R_{j\alpha}} \right| \varphi_n \right> =
\left[  \left< \frac{\partial^2 \varphi_n}{\partial r_i \partial r_j} \left| \hat{v}_{\alpha} \right| \varphi_n \right>\right.\\ + 
 \left. \left< \frac{\partial \varphi_n}{\partial r_i} \left| \hat{v}_{\alpha}\right| \frac{\partial \varphi_n}{\partial r_j} \right>\right] + {\rm c.c.}\ .
\end{multline}

%When performing geometry optimization on real-space grids, the spacing should be one of the most important aspects to consider in order to get accurate results. However, one of the weaknesses in previous versions of Octopus was the prediction of low-energy structures, performing geometry optimizations, regardless of the spacing of the mesh. In low-dimensional systems, this is a significant issue, due the precise determination of the geometric arrangement of the atoms, is probably the first step in their description and understanding~\cite{Goedecker2005}.

With our approach, the forces tend to converge faster with the grid spacing than the energy. This means that to perform geometry optimizations it would be ideal to have a local minimization method that only relies on the forces, without needing to evaluate the energy, as both values will not be entirely consistent. Such a method is the fast inertial relaxation engine (FIRE) algorithm, put forward by Bitzek {\it et al.}~\cite{Bitzek2006}. FIRE has shown a competitive performance compared with both the standard conjugate-gradient method, and more sophisticated variations typically used in {\it ab initio} calculations. A recent article shows also the FIRE as one of the most convenient algorithm due to its speed and precision to reach the nearest local minimum starting from a given initial configuration~\cite{Asenjo2013}. 

The FIRE algorithm is based on molecular dynamics with additional velocity modifications and adaptive time steps which only requires first derivatives of the target function. In the FIRE algorithm, the system slides down the potential-energy surface, gathering ``momentum'' until the direction of the gradient changes, at which point it stops, resets the adaptive parameters, and resumes sliding. This gain of momentum is done through the modification of the time step $\Delta t$ as adaptive parameter, and by introducing the following velocity modification
\begin{equation}
  \label{eq:velocity_fire}
\mathrm{\vec{v}}(t) \rightarrow \mathrm{\vec{V}}(t) = (1-\alpha)\mathrm{\vec{v}}(t) + \alpha \left| \mathrm{\vec{v}}(t) \right| \hat{F}(t)\ ,
\end{equation}
where $\mathrm{\vec{v}}$ is the velocity of the atoms, $\alpha$ is an adaptive parameter, and $\hat{F}$ is a unitary vector in the direction of the force $\vec{F}$. By doing this velocity modification, the acceleration of the atoms is given by
\begin{equation}
  \label{eq:acceleration_fire}
\dot{\mathrm{\vec{v}}}{(t)} = \dfrac{\vec{F}{(t)}}{m} - \dfrac{\alpha}{\Delta t} \left| \mathrm{\vec{v}}(t) \right| \left[\hat{\mathrm{\vec{v}}}(t)-\hat{F}(t)\right]\ ,
\end{equation}
where the second term is an introduced acceleration in a direction ``steeper'' than the usual direction of motion. Obviously, if $\alpha = 0$ then $\mathrm{\vec{V}}(t) = \mathrm{\vec{v}}(t)$, meaning the velocity modification vanishes, and the acceleration $\dot{\mathrm{\vec{v}}}{(t)} = \vec{F}{(t)}/m$, as usual.

\begin{figure}[h!]
\begin{center}
\includegraphics[width=0.9\columnwidth]{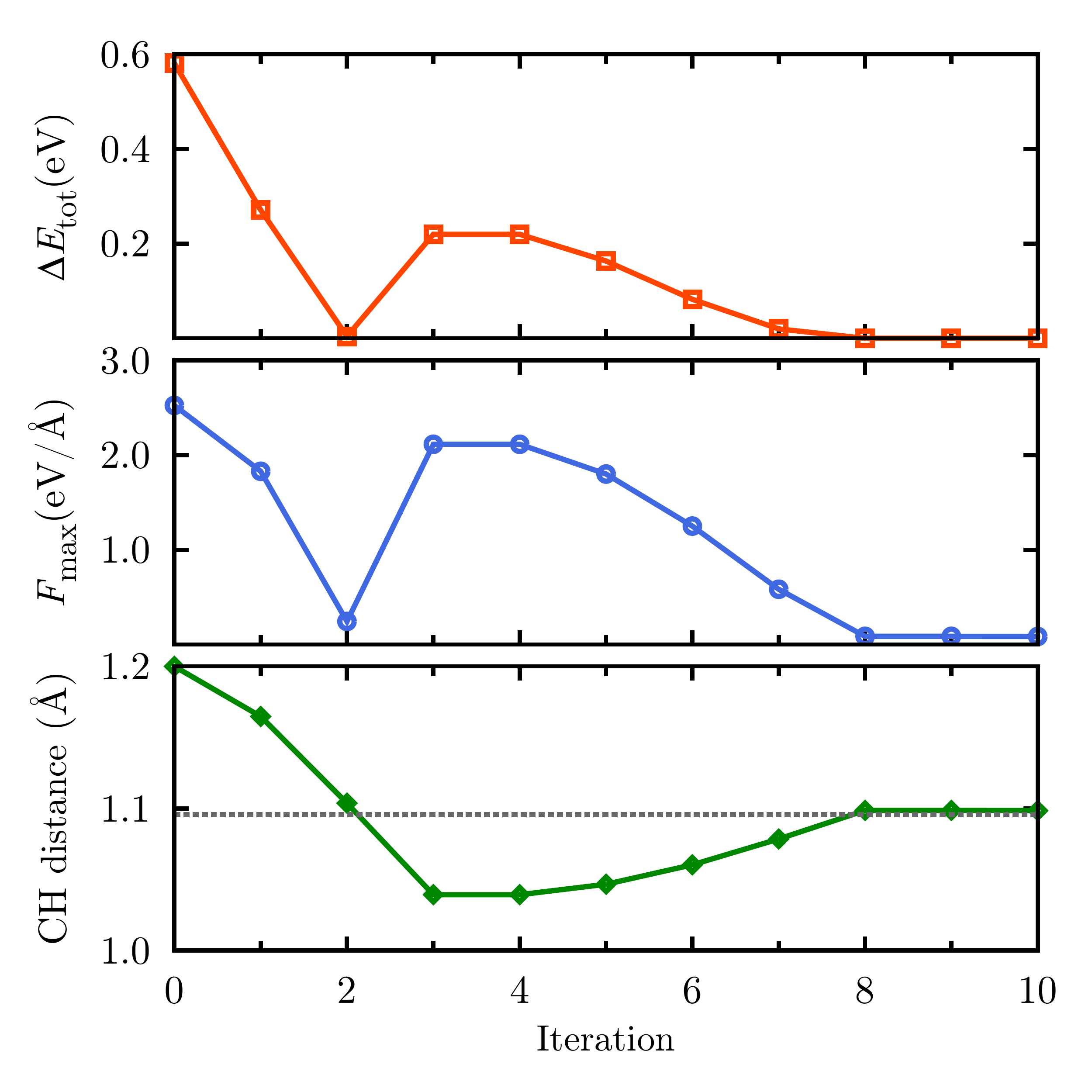}
\caption{\label{fig:go_fire}
Geometry optimization of a methane molecule with FIRE. Top panel (orange squares): energy difference $\Delta E_{\text{tot}}$ with respect to the equilibrium geometry. Middle panel (blue circles): maximum component of the force $F_{\text{max}}$ acting on the ions. Bottom panel (green diamonds): C-H bond length. Grid spacing is 0.33~Bohr.}
\end{center}
\end{figure}

We illustrate how the algorithm works with a simple case: the geometry optimization of a methane molecule. The input geometry consists of one carbon atom at the center of a tetrahedron, and four hydrogen atoms at the vertices, where the initial C-H distance is 1.2~\AA. In Fig.~\ref{fig:go_fire} we plot the energy
difference $\Delta E_{\text{tot}}$ with respect to the equilibrium conformation, the maximum component of the force acting on the ions $F_{\text{max}}$, and the C-H bond length. On the first iterations, the geometry approaches the equilibrium position, but moves away on the 3rd. This means a change in the direction of the gradient, so there is no movement in the 4th iteration, the adaptive parameters are reset, and sliding resumes in the 5th iteration.

\section{Photoemission}
Electron photoemission embraces all the processes where an 
atom, a molecule or a surface is ionized under the effect of an external electromagnetic field.
In experiments, the ejected electrons are measured with detectors that are capable 
of characterizing their kinetic properties. 
Energy-resolved, $P(E)$, and momentum-resolved, $P(\vec{k})$, photoemission probabilities are quite 
interesting observables since they carry important information, for instance, on the parent 
ion~\cite{Puschnig_2009,Wie_ner_2014} or on the ionization process itself~\cite{Huismans_2010}. 
The calculation of these quantities is a difficult task because the process requires the 
evaluation of the total wavefunction in an extremely large portion of space (in principle a macroscopic one) 
that would be impractical to represent in real space. 

We have developed a scheme to calculate photoemission based on real-time TDDFT that is currently implemented in Octopus. We use a mixed real- and momentum-space approach. Each KS orbital is propagated in real space on a restricted simulation box, and then 
matched at the boundary with a momentum-space representation.

\begin{figure}[h!]
\begin{center}
\includegraphics[width=0.9\columnwidth]{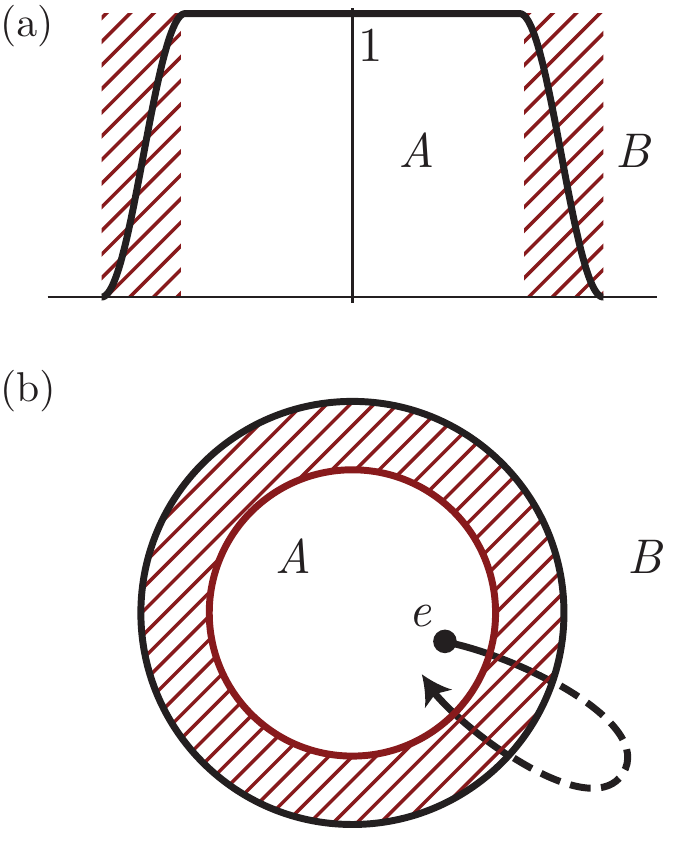}
\caption{\label{fig:pes_sheme}
Scheme illustrating the mask method for the calculation of electron photoemission.
A mask function (a) is used to effectively split each Kohn-Sham orbital into bounded and unbounded
components localized in different spatial regions $A$ and $B$ according to the diagram in (b). 
In $A$ the states are represented on a real-space grid while in $B$ they are described in momentum space.
A striped region indicates the volume where the two representations overlap.
The propagation scheme of eqs.~\eqref{eq:FMM_prop} and~\eqref{eq:FMM_prop_aux} allows seamless transitions from one representations to the other and
is capable to describe electrons following closed trajectories like the one in (b).}
\end{center}
\end{figure}

The matching is made with the help of a mask function $M(\vec{r})$, like the one shown 
in Fig.~\ref{fig:pes_sheme}, that separates each orbital into a bounded $\phi_i^A(\vec{r})$ 
and an unbounded component $\phi_i^B(\vec{r})$ as follows:
\begin{equation}\label{eq:mask_split}
  \phi_i(\vec{r},t) = \underbrace{M(\vec{r})\,\phi_i(\vec{r},t)}_{\phi_i^A(\vec{r},t)}+\underbrace{\left[1-M(\vec{r})\right]\phi_i(\vec{r},t)}_{\phi_i^B(\vec{r},t)}\, .
\end{equation}

Starting from a set of orbitals localized in $A$ at $t=0$ it is possible to derive a 
time-propagation scheme with time step $\Delta t$ by recursively applying 
the discrete time-evolution operator $\hat{U}(\Delta t)\equiv \hat{U}(t+\Delta t,t)$ and splitting  
the components with eq.~\eqref{eq:mask_split}.
The result can be written in a closed form for $\phi^A_i(\vec{r},t)$, represented in real space, and 
$\phi^B_i(\vec{k},t)$, in momentum space, with the following structure:
\begin{align}\label{eq:FMM_prop}
    \begin{array}{l}
      \phi^A_i(\vec{r},t+\Delta t) = 
      \varphi^A_i(\vec{r},t+\Delta t) +\varphi^B_i(\vec{r},t+\Delta t)\ ,\\
      \phi^B_i(\vec{k},t+\Delta t) =
      \vartheta^A_i(\vec{k},t+\Delta t)+\vartheta^B_i(\vec{k},t+\Delta t)\ ,
    \end{array}
\end{align}
and the additional set of equations,
\begin{align}\label{eq:FMM_prop_aux}
  \varphi^A_i(\vec{r},&t+\Delta t) =   M \hat{U}(\Delta t) \phi^A_i(\vec{r},t)\ ,\\
  \varphi^B_i(\vec{r},&t+\Delta t) =
  \frac{M}{(2\pi)^{3/2}}\int {\rm d}\vec{k} \mathrm{e}^{\mathrm{i}\vec{k}\cdot\vec{r}} \hat{U}_{\rm v}(\Delta t)
   \phi^B_i(\vec{k},t) \ ,\\
  \vartheta^A_i(\vec{k},&t+\Delta t) = \nonumber \\
   &\ \frac1{(2\pi)^{3/2}} \int  {\rm d}\vec{r} \mathrm{e}^{-\mathrm{i}\vec{k}\cdot\vec{r}} (1-M) \hat{U}(\Delta t)
  \phi^A_i(\vec{r},t) \ ,\\
  \vartheta^B_i(\vec{k},&t+\Delta t) =
  \hat{U}_{\rm v}(\Delta t) \phi^B_i(\vec{k},t) \nonumber \\
  & \quad\quad\quad - \frac1{(2\pi)^{3/2}} \int  {\rm d}\vec{r} \mathrm{e}^{-\mathrm{i}\vec{k}\cdot\vec{r}}
   \varphi^B_i(\vec{r},t+\Delta t)\ .
\end{align}
The momentum-resolved photoelectron probability is then obtained directly from 
the momentum components as~\cite{DeGiovannini_2012}
\begin{equation}
  P(\vec{k})=\lim_{t\rightarrow \infty}\sum_i^N |\phi^B_i(\vec{k},t)|^2\,,
\end{equation}
while the energy-resolved probability follows by direct integration, 
$P(E)=\int_{E=|\vec{k}|^2/2}{\rm d}\vec{k}P(\vec{k})$. 

In eq.~\eqref{eq:FMM_prop_aux} we introduced the Volkov propagator $\hat{U}_{\rm v}(\Delta t)$ for the wavefunctions
in $B$. 
It is the time-evolution operator associated with the Hamiltonian $\hat{H}_{\rm v}$ describing free electrons in 
an oscillating field. 
Given a time dependent vector field $\vec{{A}}(t)$, the Hamiltonian 
$\hat{H}_{\rm v}=\frac{1}{2}\left(-i\vec{\nabla}-\frac{\vec{{A}}(t)}{c}\right)^2$ expressed in the velocity gauge is diagonal in 
momentum and can be naturally applied to $\phi^B_i(\vec{k},t)$.

For all systems that can be described by a Hamiltonian such that $\hat{H}(\vec{r},t)=\hat{H}_{\rm v}(\vec{r},t)$ 
for $\vec{r} \in B$ and all time $t$, eqs. 
\eqref{eq:FMM_prop} and \eqref{eq:FMM_prop_aux} are equivalent to a time propagation in the entire space 
$A\cup B$.
In particular, it exactly describes situations where the electrons follow trajectories crossing the boundary separating $A$ and $B$ as illustrated in Fig.~\ref{fig:pes_sheme}(b).

\begin{figure}[h!]
\begin{center}
\includegraphics[width=0.9\columnwidth]{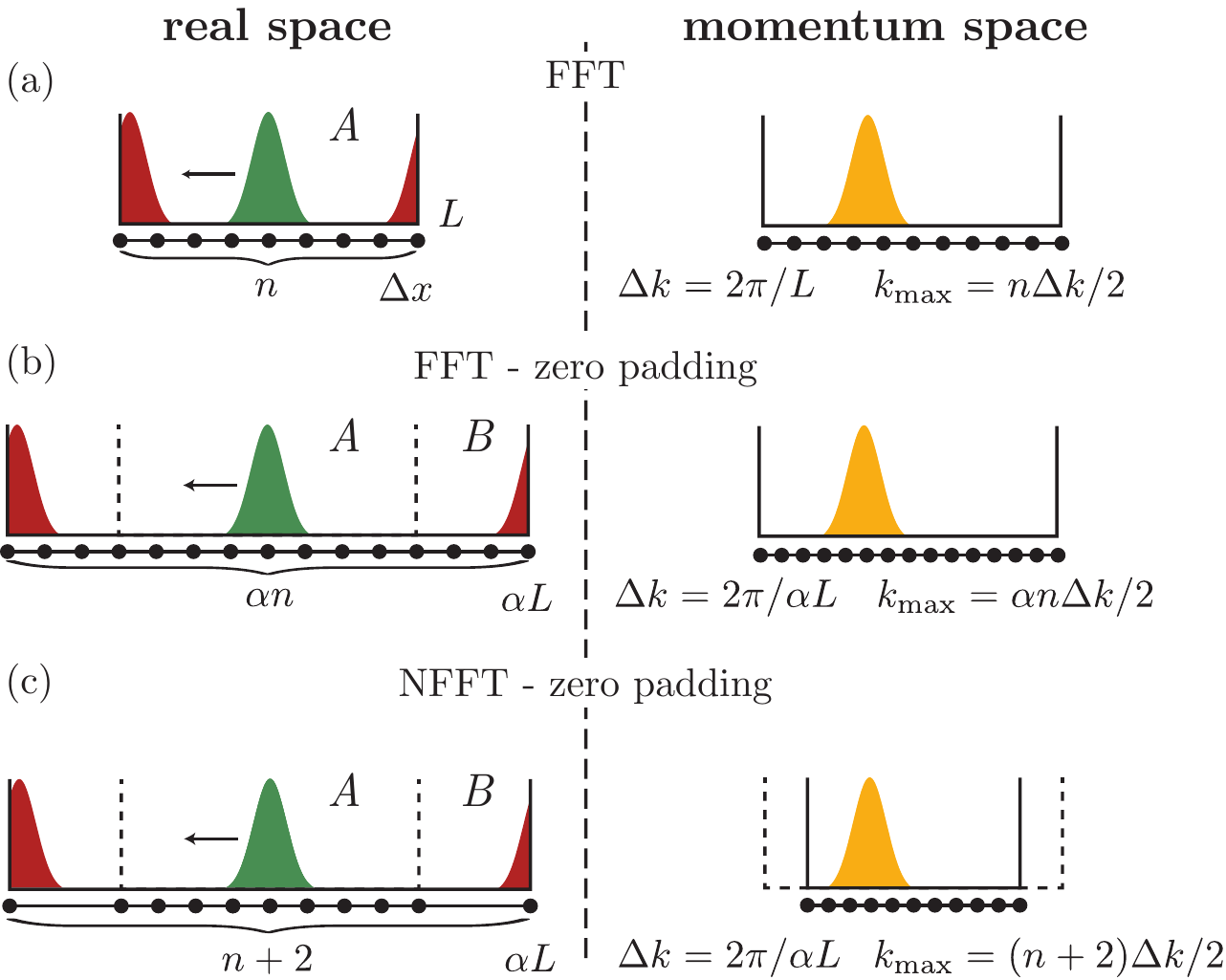}
\caption{\label{fig:pes_nfft}
Scheme illustrating different discretization strategies for eq.~\eqref{eq:FMM_prop_aux} in one dimension.
In all the cases an initial wavepacket (green) is launched towards the left side of a simulation box of 
length $L$ and discretized in $n$ sampling points spaced by $\Delta x$. $A$ and $B$ indicate the 
space partitions corresponding to Fig.~\ref{fig:pes_sheme}.
Owing to the discretization of the Fourier integrals, periodic conditions are imposed at the boundaries and
the wavepacket wraps around the edges of the simulation box (red). 
The time evolution is portrayed together with a momentum-space representation (yellow), with spacing $\Delta k$
and maximum momentum $k_{\rm max}$, in three situations differing in the strategy used to map real and
momentum spaces:(a) Fast Fourier Transform (FFT), (b) FFT extended with zeros (zero padding) in a box enlarged 
by a factor $\alpha$, and (c) zero padding with NFFT. }
\end{center}
\end{figure}

In Octopus we discretize eq.~\eqref{eq:FMM_prop_aux} in real and momentum space and co-propagate 
the complete set of orbitals $\phi^A_i(\vec{r},t)$ and $\phi^B_i(\vec{k},t)$.
The propagation has to take care of additional details since the discretization can introduce numerical 
instability.
In fact, substituting the Fourier integrals in \eqref{eq:FMM_prop_aux} with Fourier sums (usually evaluated with FFTs)
imposes periodic boundary conditions that spuriously reintroduces charge that was supposed to disappear.
This is illustrated with a one-dimensional example in Fig.~\ref{fig:pes_nfft}(a) where a wavepacket
launched towards the left edge of the simulation box reappears from the other edge.

An alternative discretization strategy is zero padding. 
This is done by embedding the system into a simulation box enlarged by a factor $\alpha>1$, extending 
the orbitals with zeros in the outer region as shown in Fig.~\ref{fig:pes_nfft}(b).
In this way, the periodic boundaries are pushed away from the simulation box and the wavepackets have to travel 
an additional distance $2(\alpha -1)L$ before reappearing from the other side.
In doing so, the computational cost is increased by adding $(\alpha -1)n$ points for each orbital.

This cost can be greatly reduced using a special grid with only two additional points placed at $\pm \alpha L$ 
as shown in Fig.~\ref{fig:pes_nfft}(c).
Since the new grid has non uniform spacing a non-equispaced FFT (NFFT) is used~\cite{Kunis_2006,Keiner_2009}. With this strategy, a price is paid in momentum space where the maximum momentum $k_{\rm max}$ is reduced 
by a factor $\alpha$ compared to ordinary FFT.
In Octopus we implemented all three strategies: bare FFT, zero padding with FFT and zero padding with NFFT.

All these discretization strategies are numerically stable for a propagation time 
approximately equivalent to the time that it takes for a wavepacket with the highest momentum considered to 
be reintroduced in the simulation box.
For longer times we can employ a modified set of equations. 
It can be derived from \eqref{eq:MM_prop_aux} under the assumption that the electron flow is only outgoing.
In this case we can drop the equation for $\varphi^B_i$ responsible for the ingoing flow
and obtain the set
\begin{align}\label{eq:MM_prop_aux}
  \begin{array}{l}
  \varphi^A_i(\vec{r},t+\Delta t) =  M \hat{U}(\Delta t) \phi^A_i(\vec{r},t)\ ,\\
  \varphi^B_i(\vec{r},t+\Delta t) = 0 \ ,\\
  \vartheta^A_i(\vec{k},t+\Delta t) = \frac1{
   (2\pi)^{{3}/{2}}} \int  {\rm d}\vec{r} e^{-\mathrm{i}\vec{k}\cdot\vec{r}} (1-M) \hat{U}(\Delta t)
  \phi^A_i(\vec{r},t) \ ,\\
  \vartheta^B_i(\vec{k},t+\Delta t) = \hat{U}_{\rm v}(\Delta t) \phi^B_i(\vec{k},t)\ .
  \end{array}
\end{align}
This new set of equations together with \eqref{eq:FMM_prop} lifts the periodic conditions at the 
boundaries and secures numerical stability for arbitrary long time propagations. 
A consequence of this approximation is the fact that the removal of charge is performed only in the equation for $\varphi^A_i$ by means of a multiplication by $M(\vec{r})$.
This is equivalent to the use of a mask function boundary absorber that is known to present reflections 
in an energy range that depends on $M(\vec{r})$~\cite{DeGiovannini:2014wo}.
Carefully choosing the most appropriate mask function thus becomes of key importance in order to 
obtain accurate results.

We conclude briefly summarizing some of the most important features and applications of our approach.
The method allows us to retrieve $P(\vec{k})$, the most resolved quantity available in experiments nowadays.
In addition, it is very flexible with respect to the definition of the external 
field and can operate in a wide range of situations. 
In the strong-field regime, it can handle interesting situations, for instance, 
when the electrons follow trajectories extending beyond the simulation box, or when the target system 
is a large molecule. 
This constitutes a step forward compared to the standard theoretical tools employed in the field which, in the 
large majority of cases, invoke the single-active-electron approximation.
In Ref.~\cite{DeGiovannini_2012} the code was successfully employed to study the photoelectron angular distributions
of nitrogen dimers under a strong infrared laser field.
The method can efficiently describe situations where more than one laser pulse is involved. 
This includes, for instance, time-resolved measurements where pump and probe setups are employed.
In Ref.~\cite{DeGiovannini_2013} Octopus was used to monitor the time evolution of the $\pi\rightarrow\pi^*$
transition in ethylene molecules with photoelectrons. 
The study was later extended to include the effect of moving ions at the classical level~\cite{CrawfordUranga_2014}.
Finally, we point out that our method is by no means restricted to the study of light-induced 
ionization but can be applied to characterize ionization induced by other processes, for example, ionization taking place after a proton collision. 

\section{Complex scaling and resonances}
% Commands to unify notation (vectors, constants, etc.) in this section.

\newcommand{\ee}{\mathrm e} % 2.718281828...
\newcommand{\ii}{\mathrm i} % sqrt -1
\newcommand{\xc}{\mathrm{xc}} % exchange--correlation
\newcommand{\Ha}{\mathrm H} % hartree
\newcommand{\dee}{\mathrm d} % dx, dr
\newcommand{\fdiff}[2]{\frac{\delta #1}{\delta #2}} % functional derivative

In this section we discuss the calculation of resonant electronic states by
means of the complex-scaling method, as implemented in Octopus.  By
``resonant states,'' we mean metastable electronic states
of finite systems, such as atoms or molecules, with a characteristic energy
and lifetime.

Mathematically, resonances can be defined as poles of the scattering
matrix or cross-section at complex energies~\cite{PhysRev.56.750,Hatano01022008}.
If a pole is close to the
real energy axis, it will produce a large, narrow peak in the
cross-section of scattered continuum states.
One way resonances can arise is from application of an
electric field strong enough to ionize the system through tunnelling.
Resonant states may temporarily capture incoming electrons or electrons
excited from bound states, making them important intermediate states
in many processes.

The defining characteristic of a resonant state, often called a Siegert
state~\cite{PhysRev.56.750}, is that it has an
outgoing component but not an incoming one.
They can be determined by solving the
time-independent Schr{\"{o}}dinger
equation with the boundary condition that the wavefunction must asymptotically
have the form
\begin{align}
  \psi(r) \sim \frac{\mathrm{e}^{\ii k r}}{r}\quad\textrm{as}\quad r\rightarrow\infty\ ,
\end{align}
where the momentum $k$ is complex and has a negative imaginary part.
This causes the state to diverge exponentially in space as
$r\rightarrow\infty$.  The state can further be ascribed a complex
energy, likewise with a negative imaginary part, causing it to decay
over time at every point in space uniformly.

Resonant states are not eigenstates of any Hermitian operator and 
in particular do not reside within the Hilbert space.  This precludes their direct
calculation with the standard computational methods
from DFT.  However, it turns out that a suitably chosen analytic
continuation of a Siegert state is localized, and this form can be used
to derive information from the state.
This is the idea behind the
\emph{complex-scaling method} \cite{aguilarcombes,Balslev:1971ez} where
states and operators are represented
by means of the transformation
\begin{align}
  \hat R_\theta\, \psi(\vec r) = \ee^{\ii N \theta / 2} \psi(\vec r \ee^{\ii\theta})\ ,
\end{align}
where $N$ is the number of spatial dimensions to which the scaling operation
is applied, and $\theta$ is a fixed \emph{scaling angle} which determines
the path in the complex plane along which the analytic continuation is taken.
The transformation maps the Hamiltonian to a non-Hermitian operator
$\hat H_\theta = \hat R_\theta \hat H \hat R_{-\theta}$.

The Siegert states $\psi(\vec r)$ of the original Hamiltonian are square-integrable
eigenstates $\psi_\theta(\vec r)$
of $\hat H_\theta$, and their eigenvalues $\epsilon_0 - \ii\Gamma/2$
define the energy $\epsilon_0$ and width $\Gamma$ of the
resonance~\cite{simon1973resonances,Reinhardt_1982,Ho19831}.

\begin{figure}[h!]
\begin{center}
\includegraphics[width=0.9\columnwidth]{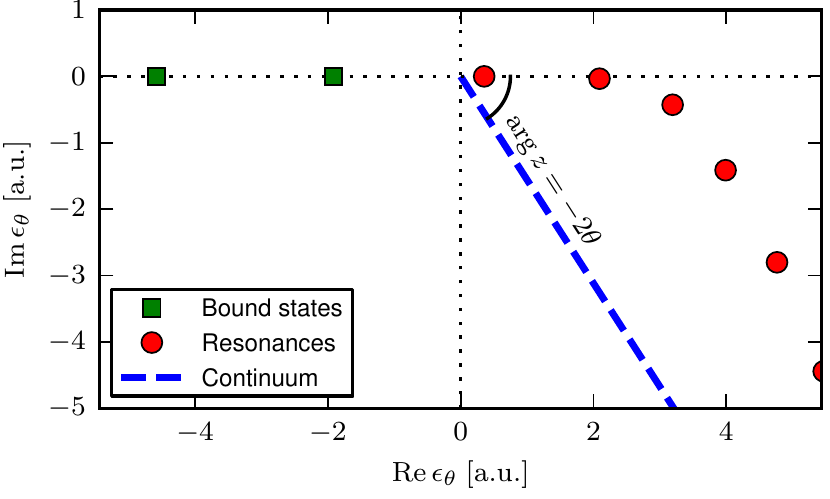}
\caption{\label{fig:cs-spectrum}
Spectrum of one-dimensional complex-scaled single-particle Hamiltonian with potential $v(x) = 3(x^2 - 2) \mathrm{e}^{-x^2 / 4}$ and $\theta=0.5$. The lowest-energy resonance, here located close to the origin, does not lie exactly on the real axis but has an imaginary part of about~$-10^{-5}$.}
\end{center}
\end{figure}

\begin{figure}[h!]
\begin{center}
\includegraphics[width=0.462\columnwidth]{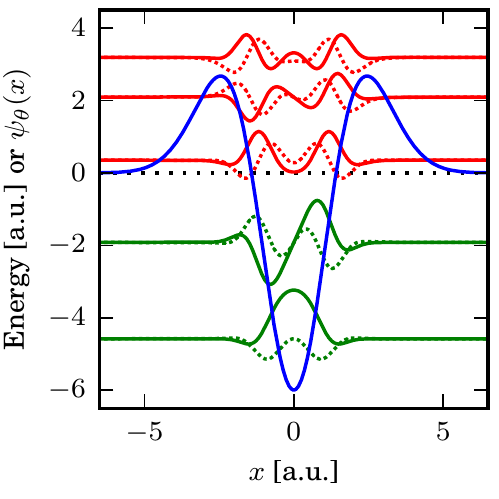}
\caption{\label{fig:cs-potential-wfs}
Potential (blue) and the real (solid) and imaginary (dotted) parts of the two bound (green) and three lowest resonant (red) wavefunctions.  For improved visualization, the wavefunctions are vertically displaced by the real parts of their energies.}
\end{center}
\end{figure}

A typical example of a spectrum of the transformed Hamiltonian $\hat H_\theta$
is shown in Fig.~\ref{fig:cs-spectrum}, and the corresponding potential
and lowest bound and resonant states in Fig.~\ref{fig:cs-potential-wfs}.
The bound-state energies are unchanged while the continuum rotates by
$-2 \theta$ around the origin.  Finally, resonances appear as isolated
eigenvalues in the fourth quadrant once $\theta$ is sufficiently large
to ``uncover'' them from the continuum.
Importantly, matrix elements (and in
particular energies) of states are independent of $\theta$ as long as
the states are localized and well represented numerically --- this
ensures that all physical
bound-state characteristics of the untransformed Hamiltonian are retained.

Our implementation supports calculations with complex scaling 
for independent particles or
in combination with DFT and
selected
xc functionals~\cite{Larsen:2013cw}.
The energy functional in KS-DFT consists of several
terms that are all expressible as integrals of the density or the
wavefunctions with the kinetic operator and various potentials.
The functional is complex-scaled as per the prescribed method
by rotating the real-space integration contour
of every term by $\theta$ in the complex plane.
The DFT energy functional becomes
\newcommand{\rprime}[0]{\vec{r}'}
\begin{multline}
  E_\theta = \ee^{-\ii2\theta}
  \sum_n \int\dee\vec r\, \varphi_{\theta n}(\vec r) \left(-\frac12 \nabla^2\right)
\varphi_{\theta n}(\vec r)\\
  + \ee^{-\ii\theta} \frac12
\iint \dee \vec r \, \dee \rprime \,
\frac{n_\theta(\vec r)n_\theta(\rprime)}{\left| \vec r - \rprime \right|}\\
\quad+ E_\xc^\theta[n_\theta]
+ \int\dee \vec r\, v_{\mathrm{ext}}(\vec r \ee^{\ii \theta}) n_\theta(\vec r)\ ,
\end{multline}
with the now-complex electron density
\begin{align}
  n_\theta(\vec r) = \sum_n f_n \varphi_{\theta n}^2(\vec r)\ ,
\end{align}
with occupation numbers $f_n$, and complex-scaled KS states $\varphi_{\theta n}(\vec r)$.
Note that no complex conjugation is performed on the left component in
matrix elements such as the density or kinetic energy.
In order to define the complex-scaled xc potential, it is necessary to perform an analytic continuation procedure~\cite{Larsen:2013cw}.

In standard DFT, the KS equations
are obtained by taking the functional derivative of the energy functional with respect
to the density.  Solving the equations corresponds to searching
for a stationary point, with the idea that this minimizes the energy.
In our case, since the energy functional is complex-valued~\cite{WM07}, we cannot minimize the energy functional, but we can still search for stationary points to find the resonances~\cite{Whitenack_2010,WW11}.
The complex-scaled version
of the KS equations thereby becomes similar to the usual ones:
\begin{align}
  \left[-\frac12 \ee^{-\ii2\theta}\nabla^2 + v_\theta(\vec r)
    \right] \varphi_{\theta n}(\vec r) = \varphi_{\theta n}(\vec r) \epsilon_{\theta n}\ .
\end{align}
The effective potential $v_\theta(\vec r)$ is the functional derivative
of the energy functional with respect to the density $n_\theta(\vec r)$, and, therefore,
consists of the terms
\begin{align}
  v_\theta(\vec r) \equiv \fdiff{E}{n_\theta(\vec r)} = 
  v_\Ha^\theta(\vec r) + v_\xc^\theta(\vec r) 
  + v_{\mathrm{ext}}(\vec r \ee^{\ii\theta})\ ,
\end{align}
where $v_{\mathrm{ext}}(\vec r \ee^{\ii\theta})$ may represent atomic
potentials as analytically
continued pseudopotentials, and where the Hartree potential
\begin{align}
  v_\Ha^\theta(\vec r) &=
  \ee^{-\ii\theta}\int\dee \rprime\, \frac{n_\theta(\vec r')}{\left| \vec r'-\vec r\right|}
\end{align}
is determined by solving the Poisson equation defined by the complex density.
Together with the xc potential, 
\begin{align}
  v_\xc^\theta(\vec r) &= \fdiff{E_\xc^\theta[n_\theta]}{n_\theta(\vec r)},
\end{align}
this defines a self-consistency cycle very similar to ordinary
KS DFT, although more care must be taken to occupy the correct states,
as they are no longer simply ordered by energy.

\begin{figure}[h!]
\begin{center}
\includegraphics[width=0.9\columnwidth]{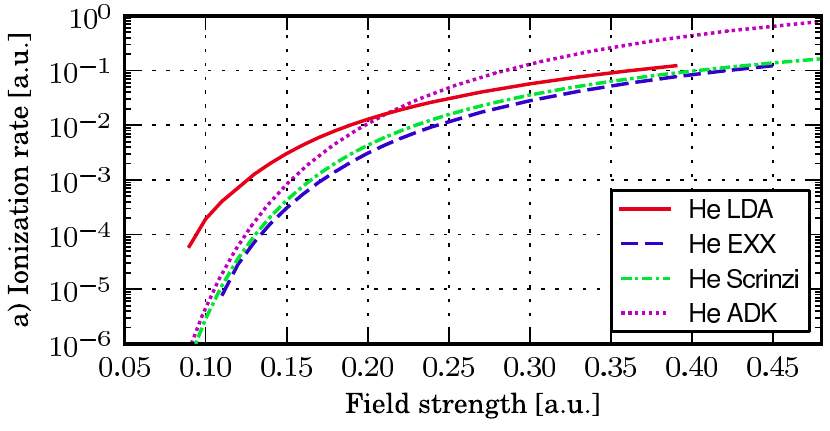}
\caption{\label{fig:cs-ionization-He} Ionization rates of the He atom in strong electric fields using the local density approximation (LDA) and exact exchange (EXX), compared to an accurate numerical reference~\cite{PhysRevLett.83.706} as well as the analytic ADK approximation~\cite{adk}.  Results from Ref.~\cite{Larsen:2013cw}}
\end{center}
\end{figure}

Fig.~\ref{fig:cs-ionization-He} shows calculated ionization rates for the He~1$s$~state in a
uniform Stark-type electric field as a function of field strength.
In the limit of weak electric fields, the simple approximation
by Ammosov, Delone and Krainov (ADK)~\cite{adk}, which depends
only on the ionization potential, approaches the accurate reference calculation
by Scrinzi and co-workers~\cite{PhysRevLett.83.706}.
This demonstrates that the ionization rate is determined largely by the
ionization potential for weak fields.  As the local density approximation is known to produce inaccurate ionization potentials
due to its wrong asymptotic form at large distances, it necessarily yields
inaccurate rates at low fields.
Meanwhile exact exchange, which is known to produce
accurate ionization energies, predicts ionization rates
much closer to the reference calculation.  The key property 
of the xc functional that allows accurate determination of decay rates
from complex-scaled DFT therefore appears to be that it must yield
accurate ionization potentials, which is linked to its ability
to reproduce the correct asymptotic form of the potential at large distances
from the system~\cite{PhysRevA.49.2421}.

\section{Quantum optimal control}

In recent years, we have added to Octopus some of the key advancements of quantum
optimal-control theory (QOCT)~\cite{Brif2010,Werschnik2007}. In this section,
we will briefly summarize what this theory is about, overview the current
status of its implementation, and describe some of the results that have been
obtained with it until now.

\textit{Quantum control} can be loosely defined as the manipulation of physical
processes at the quantum level. We are concerned here with the theoretical
branch of this discipline, whose most general formulation is precisely
QOCT. This is, in fact, a particular case of the general mathematical field of
``optimal control'', which studies the optimization of dynamical
processes in general. The first applications of optimal control in the quantum
realm appeared in the 80s~\cite{Shi1988,Peirce1988,Kosloff1989a}, and the
field has rapidly evolved since then. Broadly speaking, QOCT attempts to
answer the following question: given a quantum process governed by a
Hamiltonian that depends on a set of parameters, what are the values of those
parameters that maximize a given observable that depends on the behavior of
the system? In mathematical terms: let a set of parameters $u_1,\dots,u_M \equiv u$ determine the Hamiltonian of a system $\hat{H}[u,t]$, so that the
evolution of the system also depends on the value taken by those parameters:
\begin{align}
{\rm i}\frac{{\rm d}}{{\rm d}t}\vert\psi(t)\rangle & = 
\hat{H}[u,t]\vert\psi(t)\rangle\,,
\\
\vert\psi(0)\rangle & = \vert\psi_0\rangle\,,
\end{align}
\textit{i.e.}\ the solution of the Schr{\"{o}}dinger equation determines a map $u
\longrightarrow \psi[u]$. Suppose we wish to optimize a
functional of the system $F=F[\psi]$. QOCT is about finding the extrema of
$G(u)=F[\psi[u]]$. Beyond this search, QOCT also studies topics such
as the robustness of the optimal solutions for those parameters, the number of
solutions, or the construction of suitable algorithms to compute them.

Perhaps the most relevant result of QOCT is the equation for the gradient of $G$, which allows use of the various maximization algorithms available. For the simple formulation given above, this gradient is given by
\begin{equation}
\frac{\partial G}{\partial u_m}(u) = 2\ {\rm Im} \int_0^T\!\!{\rm d}t\;
\left< \chi(t)\left|\frac{\partial \hat{H}}{\partial u_m}[u,t]\right|
\psi(t) \right>\,,
\label{eq:qoctgradient}
\end{equation}
where $\chi$ is the \emph{costate}, an auxiliary wave function that is defined through the following equation of motion:
\begin{align}
{\rm i}\frac{\rm d}{{\rm d}t}\vert\chi(t)\rangle & =  \hat{H}^\dagger[u,t]
\vert\chi(t)\rangle\ ,
\\
\vert \chi(T) \rangle & =  \frac{\delta F}{\delta \psi^*(T)}\ .
\end{align}
This equation assumes, in order to keep this description short, that the target functional $F$ depends on the state of the system only at the final time of the propagation $T$, \textit{i.e.}\ it is a functional of $\psi(T)$. Note the presence of a boundary value equation at the final time of the propagation, as opposed to the equation of motion for the ``real'' system $\psi$, which naturally depends on an initial value condition at time zero. With these simple equations, we may already summarize what is needed from an implementation point of view in order to perform basic QOCT calculations:

The first step is the selection of the parameters $u$, that constitute the \emph{search space}. Frequently, these parameters are simply the values that the \emph{control function} (typically, the electric-field amplitude) takes at the time intervals that are used to discretize the propagation interval, \textit{i.e.} it is a ``real-time parametrization''. However, more sophisticated parametrizations allow fine-tuning of the search space, introducing constraints and penalties into the formulation.

Then, one must choose an algorithm for maximizing multi-dimensional
functions such as $G$. One possibility is the family of \emph{gradient-less} algorithms, which only require a procedure to compute the value of the function, and do not need the gradient. In this case, the previous equations are obviously not needed. One only has to propagate the system forwards in time, which is what Octopus can do best. The value of the function $G$ can then be computed from the evolution of $\psi$ obtained with this propagation, and fed into the optimization procedure. A few gradient-less algorithms are implemented in Octopus.

The most efficient optimizations can be obtained if information about the gradient is employed. In that case, we can use standard schemes, such as the family of conjugate-gradient algorithms, or the Broyden-Fletcher-Goldfarb-Shanno (BFGS) quasi-Newton scheme -- we use the implementation of these algorithms included in the GSL mathematical library~\cite{Galassi2009}. Some \textit{ad hoc} algorithms, developed explicitly for QOCT, exist. These may in some circumstances be faster than the general purpose ones. Some of those are implemented in Octopus as well~\cite{Zhu1998,Zhu1998a,Ohtsuki1999}.

In order to compute the gradient, one must implement a \emph{backwards-propagation} scheme for the costate, which does not differ from the ones used for the normal forwards propagation~\cite{Castro_2004}. Note, however, that in some cases the backwards propagation does not have the exact same simple linear form than the forwards propagation, and may include inhomogeneous or non-linear terms. The final step is the computation of the gradient from the integral given in eq.~(\ref{eq:qoctgradient}).

The formulation of QOCT we have just sketched out is quite generic; in our case the quantum systems are
those that can be modeled with Octopus (periodic systems are not supported at the moment), and the
\textit{handle} that is used to control the system is a time-dependent electric field, such as the
ones that can be used to model a laser pulse. The set of parameters \(
\left\{u\right\}_i\) define the shape of this electric field; for example, they can be the Fourier coefficients of the field amplitude.

The usual formulation of QOCT assumes the linearity of quantum mechanics. However, the time-dependent KS equations are not linear, making both the theory and the numerics more complicated. We have extended the basic theory previously described to handle the TDDFT equations, and implemented the resulting equations in Octopus~\cite{Castro2012a}.

We conclude this section by briefly describing some of the applications of the QOCT machinery included in Octopus, which can give an idea of the range of possibilities that can be attempted.  The study presented in Ref.~\cite{Rasanen2007} demonstrates the control of single-electron states in a two-dimensional semiconductor quantum-ring model. The states whose transitions are manipulated are the current-carrying states, which can be populated or de-populated with the help of circularly polarized light. 

Reference~\cite{Rasanen2008} studies double quantum dots, and shows how the electron state of these systems can be manipulated with the help of electric fields tailored by QOCT.

Another interesting application is how to tailor the shape of femtosecond laser pulses in order to obtain maximal ionization of atoms and molecules~\cite{Castro2009e}. The system chosen to demonstrate this possibility is the H$_2^+$ molecule, irradiated with short ($\approx 5$~fs) high-intensity laser pulses.

The feasibility of using the electronic current to define the target functional of the QOCT formalism is considered in Ref.~\cite{Kammerlander2011a}. 

Finally, a series of works has studied the use of optimal control for photo-chemical control: the tailoring of laser pulses to create or break selected bonds in molecules. The underlying physical model should be based on TDDFT, and on a mixed quantum/classical scheme (within Octopus, Ehrenfest molecular dynamics). Some first attempts in this area were reported in Refs~\cite{Krieger2011,Castro14}. However, these works did not consider a fully consistent optimal control theory encompassing TDDFT and Ehrenfest dynamics. This theory has been recently presented~\cite{Castro2014}, and the first computations demonstrating its feasibility will be reported soon.

\section{Plasmonics}
\label{sec:plasmonic}

The scope of real-space real-time approaches is not confined to the
atomistic description of matter. For instance, finite-difference 
time-domain~\cite{Taflove_1980} (FDTD) is a standard numerical tool of 
computational electromagnetism, while lattice Boltzmann methods~\cite{Benzi_1992} 
(LBM) are widely used in computational fluid dynamics. Indeed, real-space
real-time approaches can be used to model physical processes on rather
different space and time scales. This observation also bears
an important suggestion: numerical methods based on real-space grids
can be used to bridge between these different space and time scales.

Numerical nanoplasmonics is a paradigmatic case for multiscale electronic-structure calculations. A nanoplasmonic system -- \textit{e.g.}, made up of metal
nanoparticles (MNPs) -- can be a few tens of nanometers across, while the
region of strong field enhancement -- \textit{e.g.}, in the gap between two
MNPs -- can be less than 1 nm across~\cite{Savage_2012}. The
field enhancement, $h\left({\bf r}\right)$, is essentially a classical
observable, defined as
\begin{equation}
h\left({\bf r}\right)=\sqrt{\frac{\left\langle {\bf E}_{\rm tot}^{2}\left({\bf r}\right)\right\rangle }{\left\langle {\bf E}_{\rm ext}^{2}\left({\bf r}\right)\right\rangle }}\;,
\end{equation}
where ${\bf E}_{\rm tot}$ is the total electric field, ${\bf E}_{\rm ext}$
is the external (or driving) electric field, and $\left\langle \cdots\right\rangle $
indicates a time average. 
Large field enhancements are the key to single molecule surface-enhanced
Raman spectroscopy (SERS) and values as large as $h>100$ (the
intensity of the SERS signal scales as $h^{4}$) are predicted by
classical electromagnetic calculations~\cite{Kneipp_2002}.

In classical calculations, the electronic response is modeled by the
macroscopic permittivity of the material. The classical Drude model
gives the following simple and robust approximation of the metal (complex)
permittivity:
\begin{equation}\label{eq:local-optics}
\epsilon_{r}\left(\omega\right)=\epsilon_{\infty}-\frac{\omega_{p}^{2}}{\omega\left(\omega+i\gamma\right)}\ .
\end{equation}
For gold, typical values of the high-frequency permittivity $\epsilon_{\infty}$, the plasma frequency $\omega_{p}$, and the relaxation rate $\gamma$,
are: $\epsilon_{\infty}=9.5$, $\hbar\omega=8.95$ eV and $\hbar\gamma=69.1$
meV~\cite{Grady_2004}. A non-local correction to the Drude
model can also be included by considering the plasmon dispersion~\cite{Dobson_2000,Raza_2011}.
The metal (complex) permittivity then reads
\begin{equation}\label{eq:non-local-optics}
\epsilon_{r}\left({\bf k},\omega\right)=\epsilon_{\infty}-\frac{\omega_{p}^{2}}{\omega\left(\omega+i\gamma\right)-\beta^{2}{k}^{2}}\;.
\end{equation}
The parameter $\beta$ can be fitted to model the experimental data, 
although the value $\beta=\sqrt{3/5}\,v_{F}$, where $v_{F}$ is the Fermi 
velocity, is suggested by the Thomas-Fermi approximation.\cite{Boardman1982} 

Regardless of the level of sophistication of the permittivity model,
all classical calculations assume that electrons are strictly confined
inside the metal surfaces. This is a safe approximation
for microscopic plasmonic structures. However, at the nanoscale the
electronic delocalization (or spillout) outside the metal surfaces
becomes comparable to the smallest features of the plasmonic nanostructure,
\textit{e.g.}, to the gap between two MNPs. In this scale, the very definition
of a macroscopic permittivity is inappropriate and the electronic
response must be obtained directly from the quantum dynamics of the
electrons.

TDDFT is currently the method of choice to model the plasmonic response
of MNPs~\cite{Zuloaga_2010,Marinica_2012,Townsend_2012,Stella_2013,
Teperik_2013,Xiang_2014,Zhang_2014}, via the simplified jellium model, in which the nuclei and core electrons are described as a uniform positive charge density, and only the valence electrons are described explicitly.
Early calculations -- especially nanospheres~\cite{Marinica_2012,Esteban_2012}
-- have suggested the existence of new charge-transfer plasmonic
modes, which have been also demonstrated by pioneering experiments~\cite{Savage_2012}. In the future, as the field of quantum plasmonics
\cite{Tame_2013}
-- \textit{i.e.}, the investigation and control of the quantum properties
of plasmons -- will further develop, the demand for accurate, yet
scalable, numerical simulations to complement the experimental findings
is expected to grow. This demand represents both a challenge and an
opportunity for computational physics.

Scaling up the TDDFT@jellium method to model larger and more complex
plasmonic nanostructures is a challenge which can be addressed by
high-performance real-space real-time codes, like Octopus. The code
has been initially applied to investigate the plasmonic response of
single gold nanospheres (Wigner-Seitz radius, $r_{s}=3.0$ bohr)~
\cite{Townsend_2012}. A clear plasmonic resonance appears in the
absorption cross section -- computed by real-time propagation --
for spheres containing a large enough number of electrons ($N_{e}>100$).
A new plasmonic mode, deemed the ``quantum core plasmon'', has been
also suggested from the analysis of the absorption cross-section.
This new mode has been further characterized by probing the sphere
at its resonance frequency. Within a real-time propagation scheme,
this is simply done by including an external electric field, the ``laser pulse'', oscillating at a given frequency.

\begin{figure}[h!]
\begin{center}
\includegraphics[width=0.9\columnwidth]{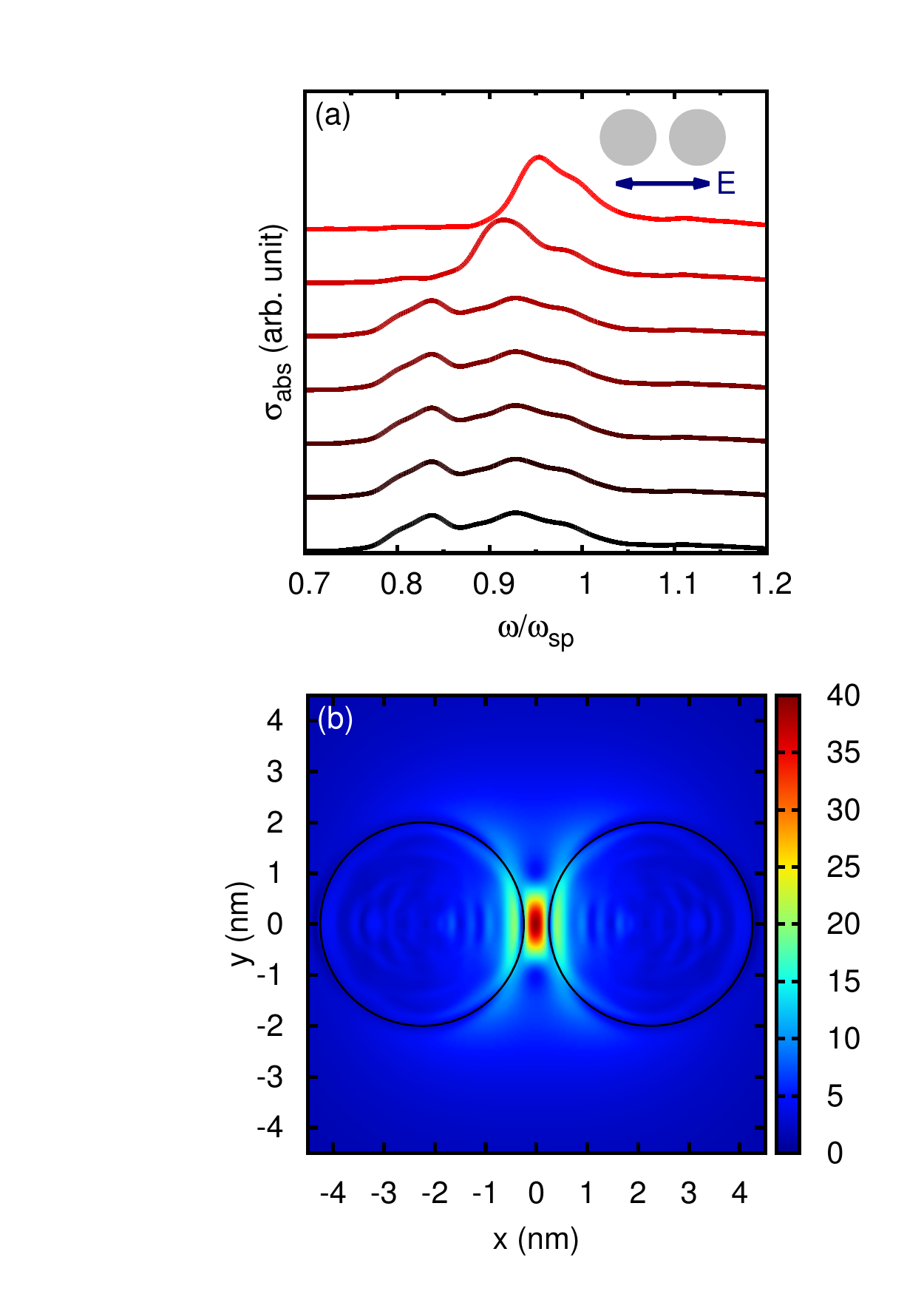}
\caption{\label{fig:FIGURE_plasmonics} Panel (a): Absorption cross section of a pair of sodium nanowires. The driving electric field is polarized as shown in the inset. Curves are for different values of gap, $d$, between the nanowires, From top to bottom: $d=5,\ 2,\ 1,\ 0.5,\ 0.2,\ 0.1,\ 0$ nm. 
Panel (b): Field enhancement, $h$, for the case $d=0.5$ nm. The black lines indicate the nanowire surfaces. (Adapted from Ref.~\cite{Stella_2013})}
\end{center}
\end{figure}

As versatility is a major strength of real-space real-time approaches,
other jellium geometries can be easily modeled by Octopus, including
periodic structures. For instance, a pair of interacting sodium nanowires
(with periodicity along their longitudinal direction) has been investigated
to assess the accuracy of classical methods based on the model permittivity
in eq.~(\ref{eq:local-optics}) and eq. (\ref{eq:non-local-optics}).
Compared to pairs of nanospheres, nanowires
display a stronger inductive interaction due to their extended geometry~\cite{Stella_2013,Teperik_2013}.
This is manifest in the absorption cross-section which already shows
a large split of the plasmonic peak for a small gap between the
wires (see Fig.~\ref{fig:FIGURE_plasmonics}(a)). Due to the electronic spillout and the symmetry
of the system, it also turns out that the largest field enhancement
is reached at the center of the gap, not on the opposing surfaces
of the nanowires as predicted by the classical methods 
(see Fig.~\ref{fig:FIGURE_plasmonics}(b)).
The maximum field enhancement estimated by the TDDFT@jellium method
is also smaller than the classical estimates. Once again, the quantum
delocalization ignored by the classical methods plays a crucial role
in ``smearing'' the singularities of the induced field, effectively
curbing the local field enhancement.

Simple jellium geometries have been implemented in Octopus and they
can be used as effective ``superatomic pseudopotentials''. The similarity
between the jellium potential and atomic pseudopotentials can be further
exploited to develop an external ``jellium pseudopotential'' generator
to be used with Octopus. In this way, a larger selection of jellium
geometries will be made available along with refined, yet scalable,
jellium approaches to include $d$ electron screening in noble metals~\cite{Rubio_1993}. Efforts in this direction are being currently made.

Finally, a word of caution about the domain of applicability of the
TDDFT@jellium method is in order. The non-uniformity of the atomic
lattice is expected to affect the absorption cross-section of small
MNPs. A careful assessment of the lattice contributions -- including
the lattice symmetry -- on the main plasmon modes of a pair of nanosphere
is available~\cite{Zhang_2014}. This last investigation further demonstrates the possibility
to bridge between atomistic and coarse-grained electronic calculations 
by means of a real-space real-time approach.

\section{Development of exchange and correlation functionals}
%Micael, Miguel, Xavier

The central quantity of the KS scheme of DFT is the xc energy $E_{\rm xc}[n]$, which describes all
non-trivial many-body effects.  Clearly, the exact form of this
quantity is unknown and it must be approximated in any practical
application of DFT. We emphasize that the accuracy of any DFT
calculation depends solely on the form of this quantity, as this is
the only real approximation in DFT (neglecting numerical
approximations that are normally controllable).

During the past 50 years, hundreds of different forms
have appeared~\cite{Scuseria2005}. They are usually arranged in
families, which have names such as generalized-gradient approximations
(GGAs), meta-GGAs, hybrid functionals, etc. In 2001, John Perdew came
up with a beautiful idea on how to illustrate these families and their
relationship~\cite{perdew:1}. He ordered the families as rungs in a
ladder that leads to the heaven of ``chemical accuracy'', which he
christened the ``Jacob's ladder'' of density-functional approximations for
the xc energy.  Every rung adds a dependency on another quantity,
thereby increasing the precision of the functional but also increasing
the numerical complexity and the computational cost.

The first three rungs of this ladder are : (i)~the
local-density approximation (LDA), where the functional has a local
dependence on the density only; (ii)~the
generalized-gradient approximation (GGA), which includes also a local
dependence on the gradient of the density \(\nabla n(\vec r)\); and
(iii)~the meta-GGA, which adds a local dependence on the Laplacian of
the density and on the kinetic-energy density. In the fourth rung we have functionals that depend on
the occupied KS orbitals, such as exact exchange or hybrid
functionals. Finally, the fifth rung adds a dependence on the virtual
KS orbitals.

Support for the first three rungs and for the local part of the hybrid
functionals in Octopus is provided through the Libxc
library~\cite{Marques_2012}. Libxc started as a spin-off project during
the initial development of Octopus. At that point, it became clear that
the task of evaluating the xc functional was completely independent
of the main structure of the code, and could therefore be transformed into
a stand-alone library. Over the years, Libxc became more and more independent of
Octopus, and is now used in a variety of DFT codes. There are
currently more than 150 xc functionals implemented in Libxc
that are available in Octopus, a number that has been increasing steadily
over the years. All of the standard functionals are included and many
of the less common ones. There is also support for LDAs and GGAs of
systems of reduced dimensionality (1D and 2D), which allow for direct
comparisons with the direct solution of the many-body Schr\"odinger
equation for model systems described in section~\ref{sec:mbse}.

Octopus also includes support for other functionals of the fourth
rung, such as exact exchange or the self-interaction correction of
Perdew and Zunger~\cite{Zunger_1980}, through the solution of the optimized effective potential equation. This can be done exactly~\cite{K_mmel_2003}, or within the Slater~\cite{Slater_1951} or Krieger-Lee-Iafrate approximations~\cite{Krieger_1990}.

Besides the functionals that are supported by Octopus, the code has
served as a platform for the testing and development of new
functionals. For example, the method described in section
\ref{sec:mbse} can be used in a straightforward way to obtain reference
data against which to benchmark the performance of a given xc
functional, for example a one-dimensional LDA~\cite{1DLDA}. In that case, both calculations, exact and approximate,
make use of the same real-space grid approach, which makes the
comparison of the results obtained with both straightforward. Despite
the obvious advantage of using exact solutions of the many-body
problem as reference data, this is often not possible and one usually
needs to resort to the more commonly used experimental or
highly-accurate quantum-chemistry data. In this case, the flexibility
of the real-space method, allowing for the calculation of many
different properties of a wide variety of systems, is again an
advantage. Octopus has therefore been used to benchmark
the performance of xc functionals whose potential has a correct
asymptotic behavior~\cite{Oliveira_2010} when calculating ionization
potentials and static polarizabilities of atoms, molecules, and
hydrogen chains. 

In this vein, Andrade and Aspuru-Guzik~\cite{Andrade_2011} proposed a method to obtain an asymptotically correct xc potential starting from any approximation. Their method is based on considering the xc potential as an electrostatic potential generated by a fictitious xc charge. In terms of this charge, the asymptotic condition is given as a simple formula that is local in real space and can be enforced by a simple procedure. The method, implemented in Octopus, was used to perform test calculations in molecules. Additionally, with this correction procedure it is possible to find accurate predictions for the derivative discontinuity and, hence, predict the fundamental gap~\cite{Mosquera_2014}.

\section{Real-space reduced density-matrix functional theory}

An alternative approach to DFT that can model electrons using a single-particle framework is reduced density matrix functional theory (RDMFT) \cite{Gilbert_1975}. Here, we present the current results of an ongoing effort to develop a real-space version of RDMFT and to implement it in the Octopus code.

Within RDMFT, the total energy of a system is given as a functional of the one-body reduced density-matrix (1-RDM)
\begin{multline}
\gamma(\vec{r},\vec{r'})=\\
N\int\cdots\int\mathrm{d}\vec{r_2}\ldots\vec{d}\vec{r}_N\,\Psi^*(\vec{r}',\vec{r}_2...\vec{r}_N)\Psi(\vec{r},\vec{r}_2...\vec{r}_N)
\end{multline}
which can be written in its spectral representation as
\begin{equation}
\gamma(\vec{r},\vec{r'})=\sum_{i=1}^{\infty}n_{i}\phi^*_i(\vec{r'})\phi_i(\vec{r}),
\end{equation}
where the natural orbitals $\phi_i(\vec{r})$ and their occupation numbers $n_i$ are the eigenfunctions and eigenvalues of the 1-RDM, respectively.

In RDMFT the total energy is given by 
\begin{multline}
\label{eqenergy}
E=-\sum_{i=1}^\infty n_i \int \mathrm{d}\vec{r}
\phi^{*}_{i}(\vec{r})\frac{\nabla^2}{2}\phi_{i}(\vec{r})\\
+\sum_{i=1}^\infty n_i \int \mathrm{d}\vec{r}\, v_{\mathrm{ext}}(\vec{r})|\phi_{i}(\vec{r})|^{2}\\
 +\frac{1}{2}\sum_{i,j=1}^\infty n_{i} n_{j}\int \mathrm{d}\vec{r} \mathrm{d}\vec{r}' \frac{|\phi_{i}(\vec{r})|^{2} |\phi_{j}(\vec{r})|^{2}}{|\vec{r}-\vec{r}'|} + E_{\rm xc}\left[\{n_{j}\},\{\phi_{j}\}\right]\ .
\end{multline}
The third term is the Hartree energy, $E_{\rm H}$, and the fourth the xc energy, $E_{\rm xc}$. As in DFT, the exact functional of RDMFT is unknown. However, the part that needs to be approximated, $E_{\rm xc}[\gamma]$, comes, contrary to DFT, only from the electron-electron interaction, as the interacting kinetic energy can be explicitly expressed in terms of $\gamma$. 
Different approximate functionals are employed and minimized with respect to the 1-RDM in order to find the ground state energy~\cite{GPB2005,ML2008,P2013}. A common approximation for $E_{\rm xc}$ is the M\"uller functional~\cite{Mueller_1984}, which has the form
\begin{multline}
E_{\rm xc}\left[\{n_j\},\{\phi_j\}\right]=\\
-\frac{1}{2}\sum_{i,j=1}^\infty \sqrt{n_{i} n_{j}}\iint \mathrm{d}\vec{r} \mathrm{d}\vec{r}' \frac{\phi_{i}^{*}(\vec{r})\phi_i(\vec{r}') \phi_{j}^{*}(\vec{r}')\phi_j(\vec{r})}{|\vec{r}-\vec{r'}|}
\end{multline}
and is the only $E_{\rm xc}$ implemented in Octopus for the moment.

For closed-shell systems, the necessary and sufficient conditions for the 1-RDM to be $N$-representable \cite{Coleman_1963}, \textit{i.e.}\ to correspond to a $N$-electron wavefunction, is that $ 0 \leq n_{i} \leq 2$ and
\begin{equation}
\sum_{i=1}^{\infty}n_{i}=N.\label{eqsumocc}
\end{equation}
Minimization of the energy functional of eq.~(\ref{eqenergy}) is performed under the $N$-representability constraints and the orthonormality requirements of the natural orbitals,
\begin{equation}
\langle \phi_{i} | \phi_{j}\rangle = \delta_{ij}. \label{eqorth}
\end{equation}
The bounds on the occupation numbers are automatically satisfied by setting $n_{i}=2\sin^2(2\pi\vartheta_i)$ and varying $\vartheta_{i}$ without constraints.  The conditions (\ref{eqsumocc}) and (\ref{eqorth}) are taken into account via Lagrange multipliers $\mu$ and $\lambda_{ij}$, respectively. Then, one can define the following functional
\begin{multline}
\Omega\left[N,\{\vartheta_i\} ,\{\phi_i(\vec{r})\}\right]= E - \mu \left(\sum_{i=1}^\infty 2\sin^2( 2\pi\vartheta_i)-N\right)\\
-\sum_{i,j=1}^\infty \lambda_{ji}\left(\langle\phi_i|\phi_j\rangle-\delta_{ij}\right)
\end{multline}
which has to be stationary with respect to variations in $\{\vartheta_i\}$, $\{\phi_i(\vec{r})\}$ and $\{\phi_i^{*}(\vec{r})\}$. In any practical calculation the infinite sums have to be truncated including only a finite number of occupation numbers and natural orbitals. However, since the occupation numbers $n_j$ decay very quickly for $j>N$, this is not problematic.

The variation of $\Omega$ is done in two steps: for a fixed set of orbitals, the energy functional is minimized with respect to occupation numbers and, accordingly, for a fixed set of occupations the energy functional is minimized with respect to variations of the orbitals until overall convergence is achieved. As a starting point we use results from a Hartree-Fock calculation and first optimize the occupation numbers. Since the correct $\mu$ is not known, it is determined via bisection: for every $\mu$ the objective functional is minimized with respect to $\vartheta_i$ until the condition (\ref{eqsumocc}) is satisfied.

Due to the dependence on the occupation numbers, the natural-orbital minimization does not lead to an eigenvalue equation like in DFT or Hartree-Fock. The implementation of the natural orbital minimization follows the method by Piris and Ugalde~\cite{Piris}. Varying $\Omega$ with respect to the orbitals for fixed occupation numbers one obtains
\begin{equation}
\lambda_{ji} = n_i\left\langle\phi_j\left|-\frac{\nabla^2}{2}+v_{\rm ext}\right|\phi_i\right\rangle
+\int \mathrm{d}\vec{r} \frac{\delta E_{\rm Hxc}}{\delta \phi_i^{*}(\vec{r})}\phi_j^{*}(\vec{r}).
\end{equation}
At the extremum, the matrix of the Lagrange multipliers must be Hermitian, \textit{i.e.} 
\begin{equation}
\lambda_{ji}-\lambda_{ij}^{*}=0\label{eqlambda}\ .
\end{equation}
Then one can define the off-diagonal elements of a Hermitian matrix $\mathbf{F}$ as:
\begin{eqnarray}
F_{ji}=\theta(i-j)(\lambda_{ji}-\lambda^{*}_{ij})+\theta(j-i)(\lambda^{*}_{ij}-\lambda_{ji}),\label{eqF}
\end{eqnarray}
where $\theta$ is the unit-step Heaviside function. We initialize the whole matrix as $F_{ji}=(\lambda_{ji}+\lambda_{ij}^{*})/2$. In every iteration we diagonalize $\mathbf{F}$, keeping the diagonal elements for the next iteration, while changing the off-diagonal ones to (\ref{eqF}). At the solution all off-diagonal elements of this matrix vanish, hence, the matrices $\mathbf{F}$ and $\gamma$ can be brought simultaneously to a diagonal form. Thus, the $\{\phi_i\}$ which are the solutions of eq.~(\ref{eqlambda}) can be found by diagonalization of $\mathbf{F}$ in an iterative manner~\cite{Piris}. The criterion to exit the natural-orbital optimization is that the difference in the total energies calculated in two successive $\mathbf{F}$ diagonalizations is smaller than a threshold.  Overall convergence is achieved when the difference in the total energies in two successive occupation-number optimizations and the non-diagonal matrix elements of $\mathbf{F}$ are close to zero.

As mentioned above, one needs an initial guess for the natural orbitals both for the first step of occupation-number optimization but also for the optimization of the natural orbitals. A rather obvious choice would be the occupied and a few unoccupied orbitals resulting from a DFT or HF calculation. Unfortunately, there are unbound states among the HF/DFT unoccupied states which are a bad starting point for the weakly occupied natural orbitals. When calculated in a finite grid these orbitals are essentially the eigenstates of a particle in a box. Using the exact-exchange approximation (EXX) in an optimized-effective-potential framework results in a larger number of bound states than HF or the local density approximation (LDA) due to the EXX functional being self-interaction-free for both occupied and unoccupied orbitals. Using HF or LDA orbitals to start a RDMFT calculation, the natural orbitals do not converge to any reasonable shape, but even when starting from EXX one needs to further localize the unoccupied states.  Thus, we have found that in order to improve the starting point for our calculation we can multiply each unoccupied orbital by a set of Gaussian functions centered at the positions of the atoms. As the unbound states are initially more delocalized than the bound ones, we choose a larger exponent for them.

\begin{figure}[h!]
\begin{center}
\includegraphics[width=0.9\columnwidth]{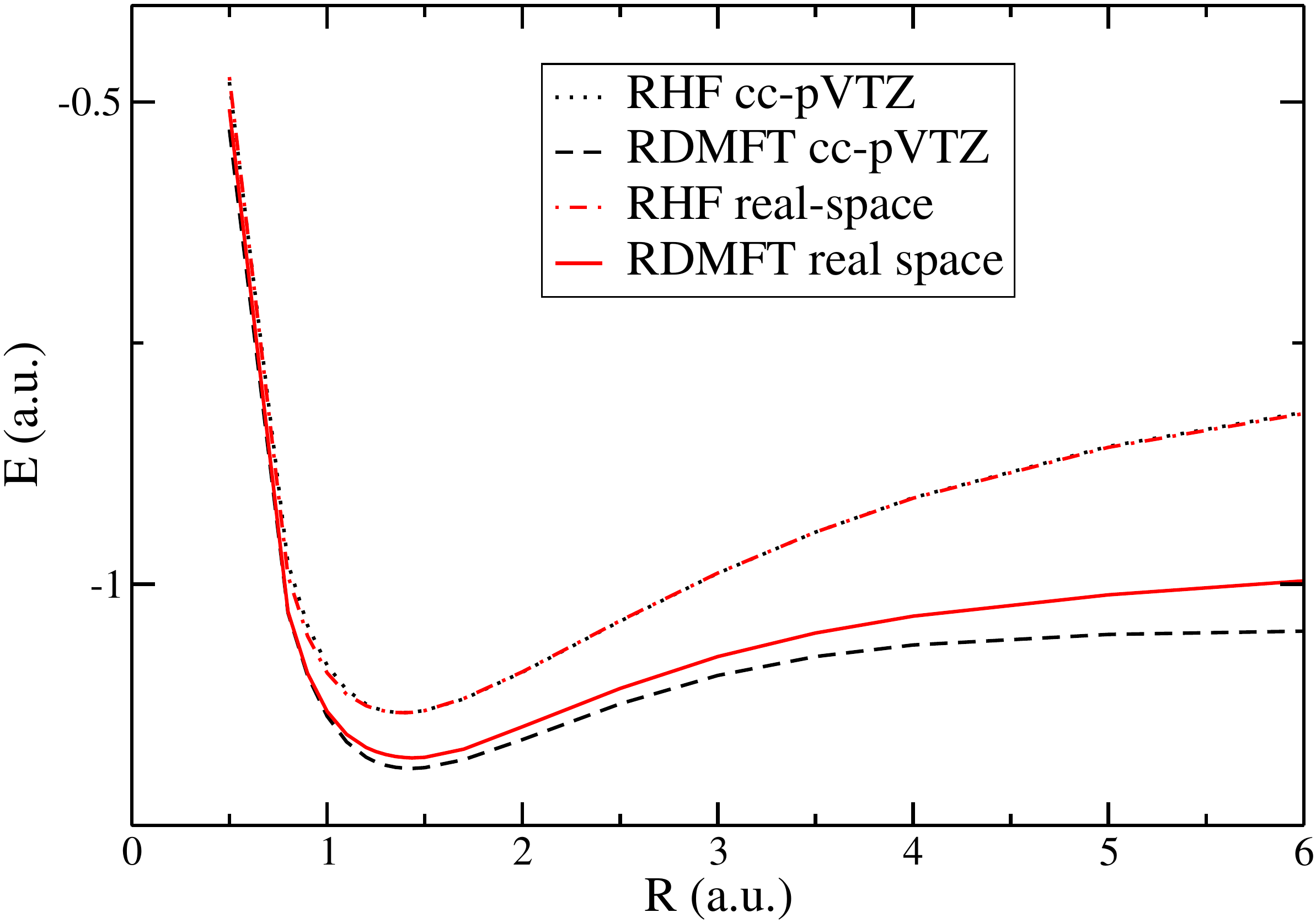}
\caption{\label{fig:rdmft_h2} Dissociation curve of the hydrogen molecule. Restricted Hartree-Fock (black dotted and red dash-dotted lines) does not dissociate into two neutral atoms while the closed-shell RDMFT gives almost the correct energy of -1 Ha at the dissociation limit in a Gaussian implementation. For the grid implementation in Octopus, a deviation from the constant energy at large $R$ remains.}
\end{center}
\end{figure}

In Fig.~\ref{fig:rdmft_h2} we show the dissociation curve of H$_{2}$ obtained with RDMFT in Octopus and compare it with results obtained by the Gaussian-basis-set RDMFT code HIPPO~\cite{ML_Benchmark}. For the Octopus calculation, we kept 13 natural orbitals with the smallest occupation number being of the order of $10^{-5}$ after the RDMFT calculation had converged. The HIPPO calculation was performed using $30$ natural orbitals. The RDMFT curve obtained with Octopus looks similar to the one from HIPPO and other Gaussian implementations of RDMFT \cite{GPB2005}, keeping the nice feature of not diverging strongly in the dissociation limit. However, for internuclear distances $R$ bigger than 1 a.u., the real-space energy lies above the HIPPO one. We believe that the remaining difference can be removed by further improving the initial guess for the orbitals that we use in Octopus, because a trial calculation using HF orbitals from a Gaussian implementation showed a curve almost identical to the one from the HIPPO code (not shown in the figure). In the future, we plan to include support for open-shell systems and additional xc functionals. 

\section{Exact solution of the many-body Schr\"odinger equation for few electrons}
\label{sec:mbse}

In one-dimensional systems, the fully interacting Hamiltonian for $N$ electrons has the form
\begin{equation}
\label{eq:1dham}
\hat{H}=\sum_{j=1}^N \left(-\frac{d^2}{dx_j^2}+v_{\rm ext}(x_j)\right)+\sum_{j<k}^N v_{\rm int}(x_j, x_k),
\end{equation}
where the interaction potential $v_{\rm int}(x_j, x_k)$ is usually Coulombic, though the following discussion also applies for other types of interaction, including more than two-body ones. In 1D one often uses the soft Coulomb interaction $1 / \sqrt{(x_j-x_k)^2+1}$, where a softening parameter (usually set to one) is introduced in order to avoid the divergence at $x_j=x_k$, which is non-integrable in 1D. 

Mathematically, the Hamiltonian (eq. \eqref{eq:1dham}) is equivalent to that of a single (and hence truly independent) electron in $N$ dimensions, with the external potential 
\begin{equation}
v_{\rm ext}^{Nd}(x_1...x_N)=\sum_{j=1}^N v_{\rm ext}(x_j)+\sum_{j<k}^N v_{\rm int}(x_j, x_k).
\end{equation}
For small $N$ it is numerically feasible to solve the $N$-dimensional Schr\"odinger equation
\begin{equation}
\label{eq:SENd}
\hat{H}\Psi_j(x_1...x_N)=E_j\Psi_j(x_1...x_N)
\end{equation}
which provides a spatial wave function for a single particle in $N$ dimensions. This equivalence is not restricted to one-dimensional problems. One can generally map a problem of $N$ electrons in $d$ dimensions onto the problem of a single particle in $Nd$ dimensions, or indeed a problem with multiple types of particles (\textit{e.g.} electrons and protons) in $d$ dimensions, in the same way. 

What we exploit in Octopus is the basic machinery for solving the Sch\"odinger equation in an arbitrary dimension, the spatial/grid bookkeeping, the ability to represent an arbitrary external potential, and the intrinsic parallelization. In order to keep our notation relatively simple, we will continue to discuss the case of an originally one-dimensional problem with $N$ electrons. Grid-based solutions of the full Schr\"odinger equation are not new, and have been performed for many problems with either few electrons (in particular H$_2$, D$_2$ and H$_2^+$)~\cite{Ranitovic21012014,lein2002}) or model interactions~\cite{luo2013}, including time-dependent cases~\cite{ramsden2012}.

The time-dependent propagation of the Schr\"odinger equation can be carried out in the same spirit, since the Hamiltonian is given explicitly and each ``single-particle orbital'' represents a full state of the system. A laser or electric-field perturbation can also be applied, depending on the charge of each particle (given in the input), and taking care to apply the same effective field to each particle along the polarization direction of the field (in 1D, the diagonal of the hyper-cube). 

Solving eq.\ (\ref{eq:SENd}) leaves the problem of constructing a wave function which satisfies the anti-symmetry properties of $N$ electrons in one dimension. For fermions one needs to ensure that those spatial wave functions $\Psi_j$ which are not the spatial part of a properly anti-symmetric wave function are removed as allowed solutions for the $N$-electron problem. A graphical representation of which wave functions are allowed is given by the Young diagrams (or tableaux) for permutation symmetries, where each electron is assigned a box, and the boxes are then stacked in columns and rows (for details see, for example, Ref.~\cite{McWeeny}). Each box is labeled with a number from 1 to $N$ such that the numbers increase from top to bottom and left to right. 
%%  NB: this is basic quantum mechanics, possibly not the place to add it here. I recommend we chop it out and add a reference to some QM textbook, even if we feel that it is not simple or well explained anywhere. A summary of the operation and the output of octopus would be sufficient.

\begin{figure}[h!]
\begin{center}
\includegraphics[width=0.9\columnwidth]{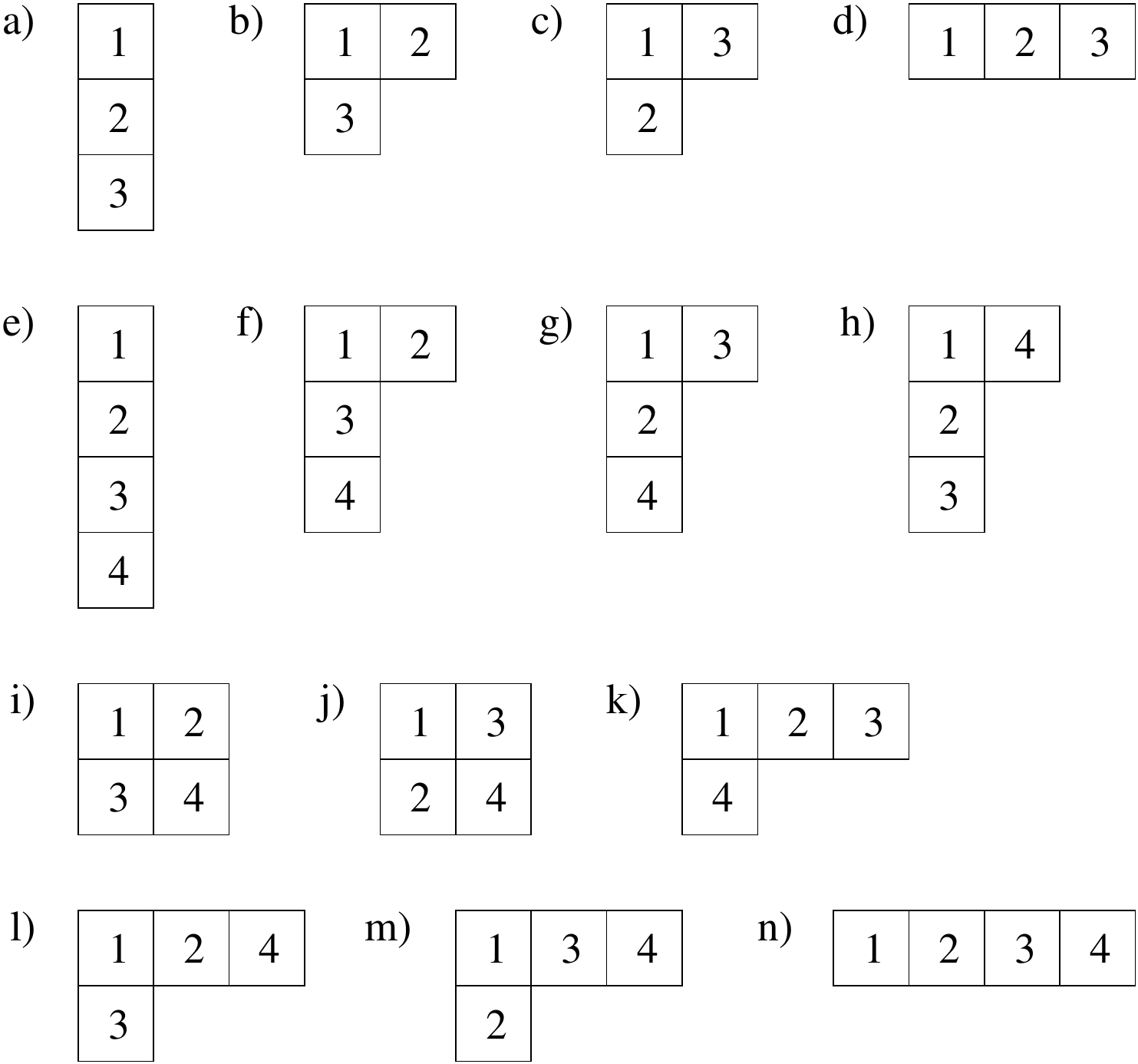}
\caption{\label{fig:young}
Young diagrams for three [a)-d)] and four [e)-n)] electrons. For three electrons, only diagrams a)-c) are allowed for spin-1/2 particles, while only diagrams e)-j) are allowed for four electrons.}
\end{center}
\end{figure}

All possible decorated Young diagrams for three and four electrons are shown in Fig.\ \ref{fig:young}. Since there are two different spin states for electrons, our Young diagrams for the allowed spatial wave functions contain at most two columns. The diagram d) is not allowed for the wave function of three particles with spin $1/2$, and diagrams k) to n) are not allowed for four particles. To connect a given wave function $\Psi_j$ with a diagram one has to symmetrize the wave function according to the diagram. For example, for diagram b) one would perform the following operations on a function \(\Psi(x_1,x_2,x_3)\)
\begin{multline}
\left[ \Psi(x_1,x_2,x_3)+\Psi(x_2,x_1,x_3)\right]\\
- \left[\Psi(x_3,x_2,x_1)+\Psi(x_3,x_1,x_2)\right].
\end{multline}
Hence, one symmetrizes with respect to an interchange of the first two variables, because they appear in the same row of the Young diagram, and anti-symmetrizes with respect to the first and third variable, as they appear on the same column. We note that we are referring to the position of the variable in the list, not the index, and that symmetrization always comes before anti-symmetrization. At the end of these operations one calculates the norm of the resulting wave function. If it passes a certain threshold, by default set to $10^{-5}$, one keeps the obtained function as a proper fermionic spatial part. If the norm is below the threshold, one continues with the next allowed diagram until either a norm larger than the threshold is found or all diagrams are used up. If a solution $\Psi_j$ does not yield a norm above the threshold for any diagram it is removed since it corresponds to a wave function with only bosonic or other non-fermionic character. Generally, as the number of forbidden diagrams increases with $N$, the number of wave functions that need to be removed also increases quickly with $N$, in particular in the lowest part of the spectrum. The case of two electrons is specific, as all solutions of eq.\ (\ref{eq:SENd}) correspond to allowed fermionic wave functions: the symmetric ones to the singlet states and the anti-symmetric ones to the triplet states.

For example, for a one-dimensional Li atom with an external potential
\begin{equation}
v_{\rm ext}(x)=-\frac{3}{\sqrt{x^2+1}}
\end{equation}
and the soft Coulomb interaction, we obtain the states and energy eigenvalues given in table~\ref{tab:young}.

\begin{table}
\centering
\begin{tabular}{r|ccr}
State & Energy & Young diagram & Norm\\ \hline
1 & -4.721 & bosonic & $<10^{-13}$\\ 
2 & -4.211 &  b) & 0.2\\
3 & -4.211 &  c) & 0.6\\
4 & -4.086 & bosonic & \(<10^{-11}\)\\
5 & -4.052 &  b) & 0.4\\
6 & -4.052 &  c) & 0.7
\end{tabular}
\caption{Eigenstates for a one-dimensional lithium atom. The first and the fourth eigenstates show norms that are smaller than $10^{-13}$ and $10^{-11}$, respectively, for all diagrams. Hence, these states are bosonic and removed from any further calculations. The second and third states are energetically degenerate and correspond to diagrams b) and c) in Fig.~\ref{fig:young}. The same is true for the fifth and sixth states.\label{tab:young}}
\end{table}

If certain state energies are degenerate, the Young diagram ``projection'' contains an additional loop, ensuring that the same diagram is not used to symmetrize successive states: this would yield the same spatial part for each wave function in the degenerate sub-space. A given diagram is only used once in the sub-space, on the first state whose projection has significant weight.

%% NB for Nicole: with a spin of 1/2 you can not use diagrams with more than 2 lines, ever - only 2 spin states exist. The number of lines gives you the different spatial orbitals. In your list above everything is ok, but the diagram below is not ok for N=5 spin 1/2 bosons:
%% 1   2   3
%% 4
%% 5
The implementation also allows for the treatment of bosons, in which case the total wave function has to be symmetric under exchange of two particles. Here one will use a spin part symmetrized with the same Young diagram (instead of the mirror one for fermions), such that the total wave function becomes symmetric. 

In order for the (anti-)symmetrization to work properly one needs to declare each particle in the calculation to be a fermion, a boson, or an anyon. In the latter case, the corresponding spatial variables are not considered at all in the (anti-)symmetrization procedure. One can also have more than one type of fermion or boson, in which case the symmetric requirements are only enforced for particles belonging to the same type.

There are also numerical constraints on the wave-functions: space must be represented in a homogeneous hyper-cube, eventually allowing for different particle masses by modifying the kinetic-energy operator for the corresponding directions. All of the grid-partitioning algorithms intrinsic to octopus carry over to arbitrary dimensions, which allows for immediate parallelization of the calculations of the ground and excited states. The code can run with an arbitrary number of dimensions, however, the complexity and memory size grow exponentially with the number of particles simulated, as expected. Production runs have been executed up to 6 or 7 dimensions.

Most of the additional treatment for many-body quantities is actually post-processing of the wave-functions. For each state, the determination of the fermionic or bosonic nature by Young-tableau symmetrization is followed by the calculation and output of the density for each given particle type, if several are present. Other properties of the many-body wave-function can also be calculated. For example, Octopus can also output the one-body density matrix, provided in terms of its occupation numbers and natural orbitals.

This type of studies, even when they are limited to model systems of a few electrons, allows us to produce results that can be compared to lower levels of theory like approximate DFT or RDMFT, and to develop better approximations for the exchange and correlation term. Exact results obtained from such calculations have been used to assess the quality of a 1D LDA functional~\cite{1DLDA} and adiabatic 1D LDA and exact exchange in a TDDFT calculation calculation of photoemission spectra~\cite{1DLDA,Helbig1}. 

\section{Compressed sensing and atomistic simulations}

In order to obtain frequency-resolved quantities from real-time methods like molecular dynamics or electron dynamics, it is necessary to perform a spectral representation of the time-resolved signal. This is a standard operation that is usually performed using a discrete Fourier transform. Since the resolution of the spectrum is given by the length of the time signal, it is interesting to look for more methods that can provide us a spectrum of similar quality with shorter time series, as this is directly reflected in shorter computation times.
Several such methods exist, but a particular one that has been explored in Octopus, due to its general applicability and efficiency, is compressed sensing.

Compressed sensing~\cite{Candes_2008} is a general theory aimed at optimizing the amount of sampling required to reconstruct a signal. It is based on the idea of sparsity, a measure of how many zero coefficients a signal has when represented in a certain basis. Compressed sensing has been applied to many problems in experimental sciences~\cite{Blumensath_2009,Zhu_2012,Sanders_2012} and technology~\cite{Schewe_2006,Harding_2013} in order to perform more accurate measurements. Its ideas, however, can also be applied to computational work.

In order to calculate a spectrum in compressed sensing, we need to solve the so-called basis-pursuit optimization problem
\begin{equation}
  \label{eq:bp}
  \min_{\vec{\sigma}} |\vec{\sigma}|_1 \quad \textrm{subject to}\quad {\mathrm{F}}\vec{\sigma}=\vec{\tau}\ ,
\end{equation}
where \(|\sigma|_1=\sum_k|\sigma_k|\) is the standard 1-norm, \(\vec\tau\) is the discretized time series, \(\vec\sigma\) is the frequency-resolved function (the spectrum that we want to calculate) and \(\mathrm{F}\) is the Fourier-transform matrix. 

Since \(\vec{\tau}\) is a short signal, its dimension is smaller than the one of \(\vec\sigma\). This implies that the linear equation \({\mathrm{F}}\vec{\sigma}=\vec{\tau}\) is under-determined and has many solutions, in this particular case, all the spectra that are compatible with our short time propagation. From all of these possible solutions, eq.~(\ref{eq:bp}) takes the one that has the smallest 1-norm, that corresponds to the solution that has the most zero coefficients. For spectra, this means we are choosing the one with the fewest frequencies, which will tend to be the physical one, as for many cases we know that the spectra is composed of a discrete number of frequencies. 

To solve eq.~(\ref{eq:bp}) numerically, we have implemented in Octopus the SPGL1 algorithm~\cite{van_den_Berg_2009}. The solution typically takes a few minutes, which is two orders of magnitude more expensive than the standard Fourier transform, but this is negligible in comparison with the cost of the time propagation.

\begin{figure}[h!]
\begin{center}
\includegraphics[width=0.9\columnwidth]{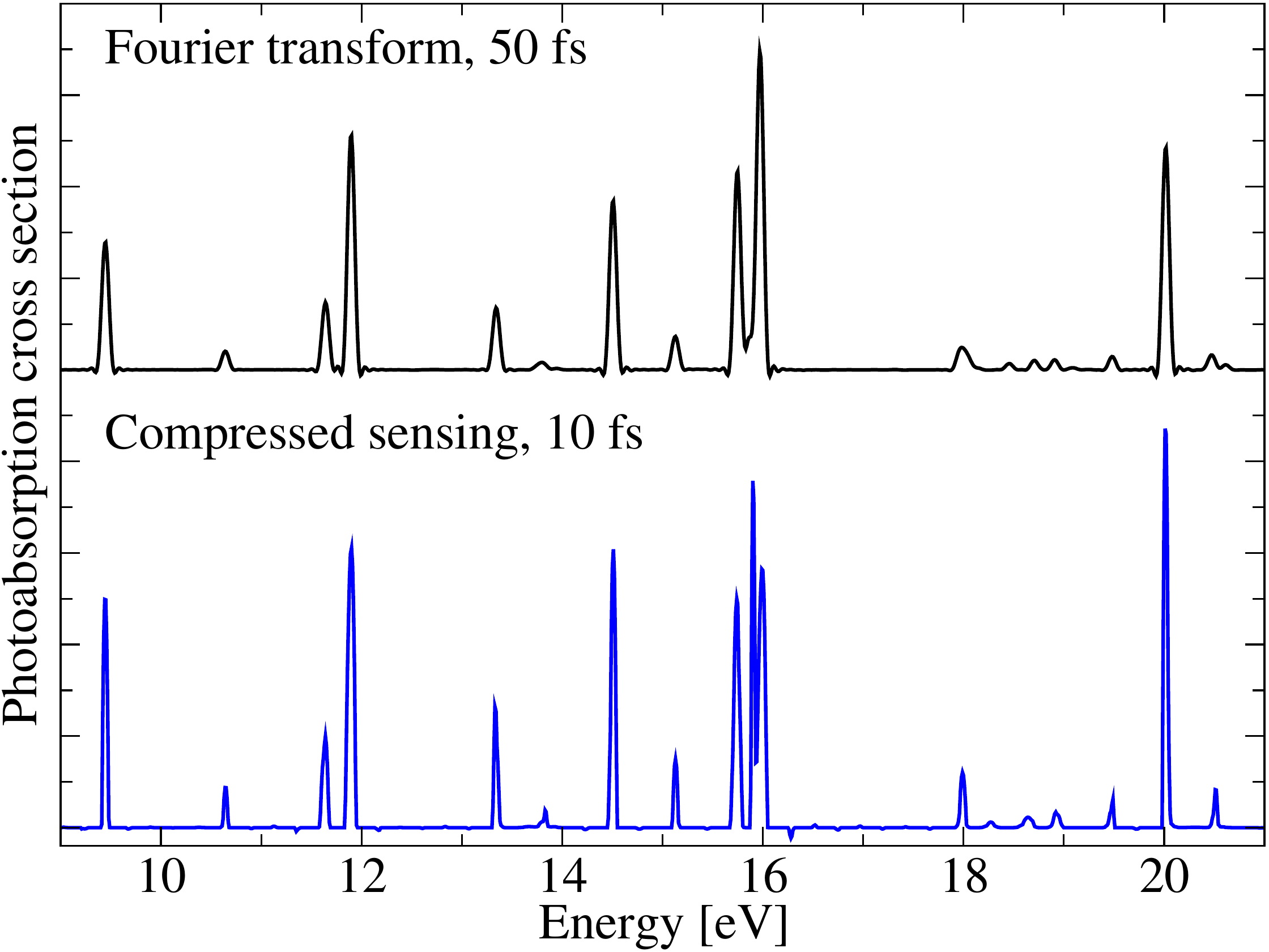}
\caption{\label{fig:cs} Optical absorption spectrum from a methane molecule from real-time TDDFT. Comparison of the calculation using a Fourier transform and a propagation time of 50 fs (top, black curve) with compressed sensing and a propagation time of 10 fs (bottom, blue curve). Compressed sensing produces a similar resolution, with a propagation 5 times shorter.}
\end{center}
\end{figure}

By applying compressed sensing to the determination of absorptional or vibrational spectra, it was found that a time signal a fifth of the length can be used in comparison with the standard Fourier transform~\cite{Andrade_2012}. This is translated into an impressive factor-of-five reduction in the computational time. This is illustrated in Fig.~\ref{fig:cs} where we show a spectrum calculated with compressed sensing from a 10 fs propagation, which has a resolution similar to a Fourier transform spectrum obtained with 50 fs of propagation time.

Moreover, the general conclusion that can be obtained from this work is that in the application of compressed sensing to simulations the reduction in the number of samples that compressed sensing produces in an experimental setup is translated into a reduction of the computational time. This concept inspired studies on how to carry the ideas of compressed sensing into the core of electronic-structure simulations. The first result of this effort is a method to use compressed sensing to reconstruct sparse matrices, that has direct application in the calculation of the Hessian matrix and vibrational frequencies from linear response (as discussed in section~\ref{sec:sternheimer}). For this case, our method results in the reduction of the computational time by a factor of three~\cite{Sanders_2015}.

\section{Parallelization, optimizations and graphics processing units}
\label{sec:parallelization}

Computational cost has been and still is a fundamental factor in the
development of electronic structure methods, as the small spatial
dimensions and the fast movement of electrons severely limit the
size of systems that can be simulated. In order to study systems of
interest as realistically and accurately as possible, electronic-structure codes must execute efficiently in modern computational
platforms. This implies support for massively parallel platforms and
modern parallel processors, including graphics processing units
(GPUs). 

Octopus has been shown to perform efficiently on parallel
supercomputers, scaling to hundreds of thousands of cores~\cite{Andrade_2012,Alberdi_2014}. The code also has an implementation of GPU acceleration~\cite{Andrade_2012_gpus,Andrade_2012} that has shown to be competitive in performance with Gaussian DFT running on GPUs~\cite{Andrade_2013}.

Performance is not only important for established methods, but also
for the implementation of new ideas. The simplicity of real-space grids allows us to provide Octopus developers with building blocks that they can use to produce
highly efficient code without needing to know the details of the
implementation, isolating them as much as possible from the optimization and parallelization requirements. In most cases, these building blocks allow developers to write
code that is automatically parallel, efficient, and that can transparently
run  on GPUs. The type of operations available run from simple ones, like
integration, linear algebra, and differential operators, to more
sophisticated ones, like the application of a Hamiltonian or solvers for differential equations.

However, it is critical to expose an interface with the adequate level
that hides the performance details, while still giving enough
flexibility to the developers. For example, we have found that the
traditional picture of a state as the basic object is not adequate for
optimal performance, as it does not expose enough data
parallelism~\cite{Andrade_2012_gpus}. In Octopus we use a higher-level
interface where the basic object is a group of several states.

In the case of functions represented on the grid, the developers work with a linear array that contains the values of
the field for each grid point. Additional data structures provide
information about the grid structure. This
level of abstraction makes it simple for developers to write code that
works for different problem dimensionality, and different kinds and
shapes of grids.

\begin{figure}[h!]
\begin{center}
\includegraphics[width=1\columnwidth]{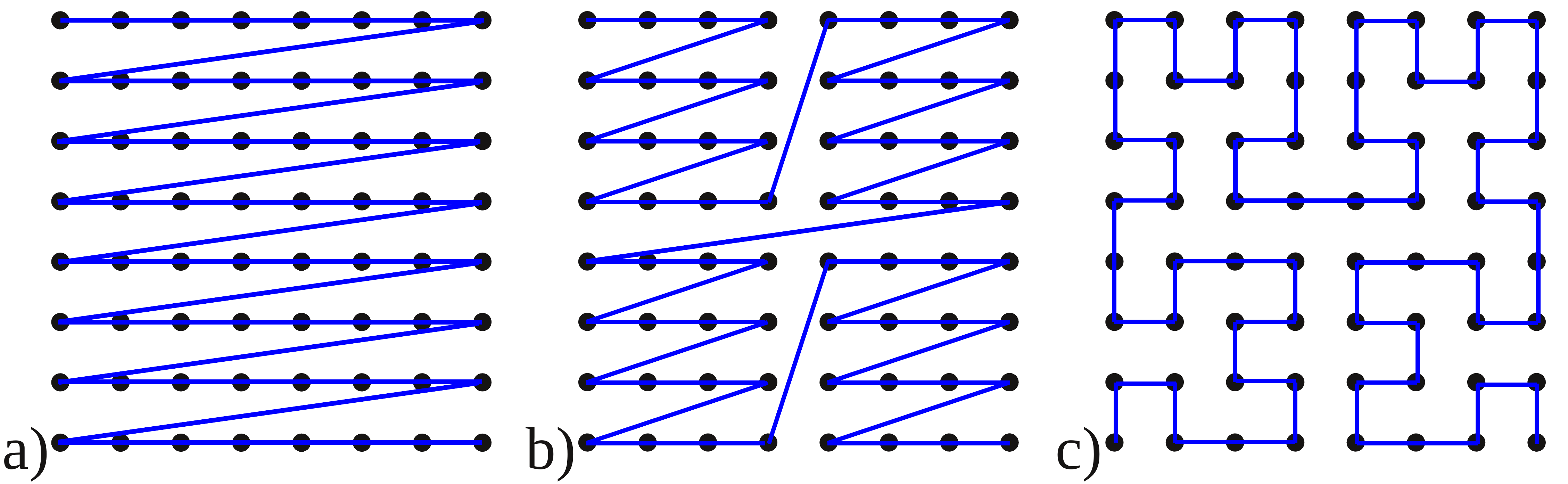}
\caption{\label{fig:gridmaps}
Examples of different mappings from a 2D grid to a linear array: (a)~standard map, (b)~grid mapped by small parallelepipedic subgrids, and (c) mapping given by a Hilbert space-filling   curve. These last two mappings provide a much better memory locality for semi-local operations than the standard approach. }
\end{center}
\end{figure}

\begin{figure}[h!]
\begin{center}
\includegraphics[width=0.9\columnwidth]{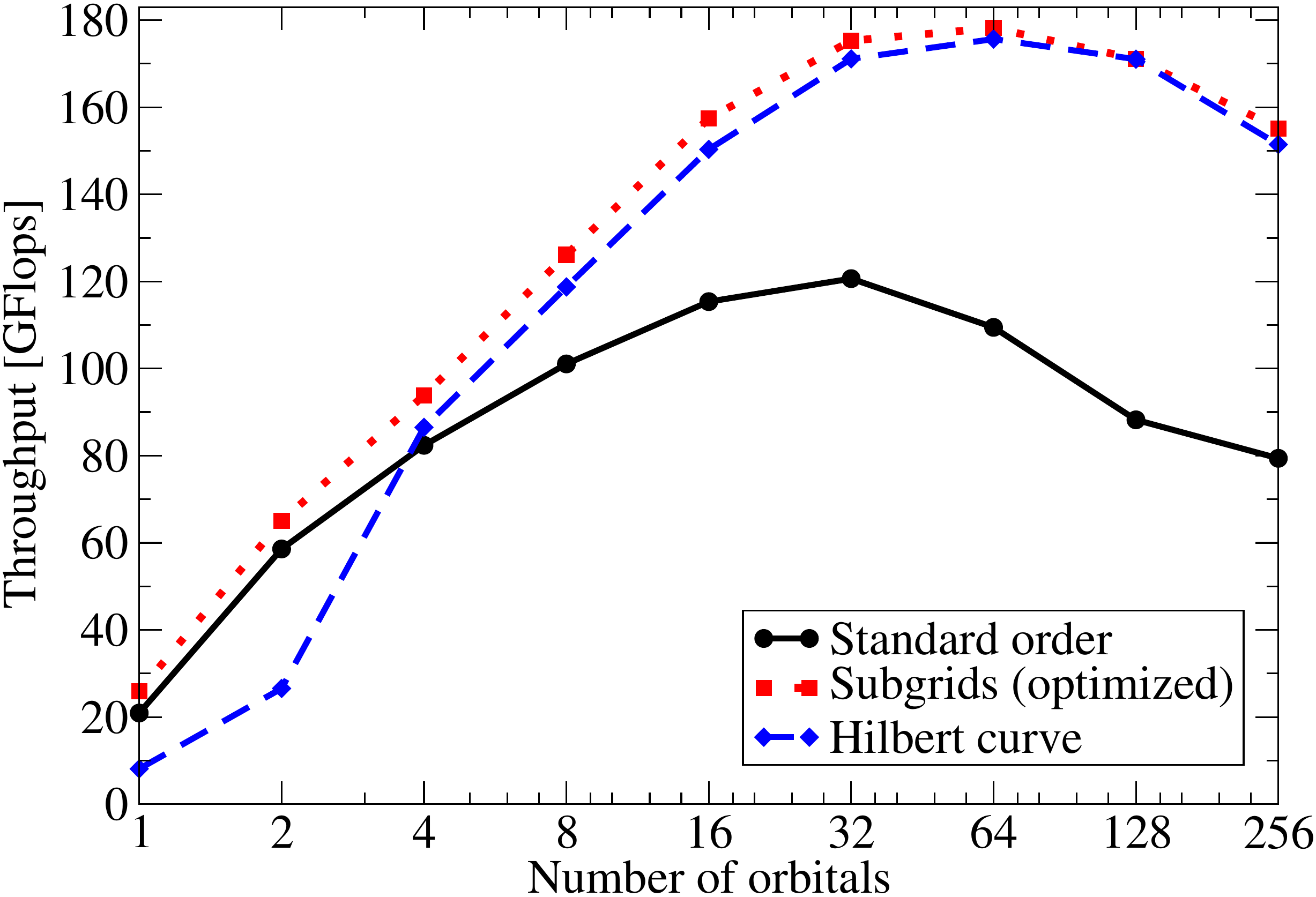}
\caption{\label{fig:gpu_laplacian}
Numerical performance of the Octopus finite-difference Laplacian implementation using different grid mappings. Spherical grid with 500,000 points. Computations with a AMD Radeon 7970 GPU. A speed up of around 50\% is observed for the subgrid and Hilbert curve mappings.}
\end{center}
\end{figure}

In terms of performance, by hiding the structure of the grid, we can
use sophisticated methods to control how the grid points are stored in
memory with the objective of using processor caches more efficiently
in finite-difference operators. We have found that by using space-filling curves~\cite{Peano_1890}, as shown in Fig.~\ref{fig:gridmaps}, and
in particular the Hilbert curve~\cite{Hilbert_1891,Skilling2004}, we can produce a significant improvement in the performance of semi-local operations. For example, in Fig.\ref{fig:gpu_laplacian} shows that a performance gain of around 50\% can be obtained for the finite-difference Laplacian operator running on a GPU by using a Hilbert curve to map the grid into memory.

\begin{figure}[h!]
\begin{center}
\includegraphics[width=1\columnwidth]{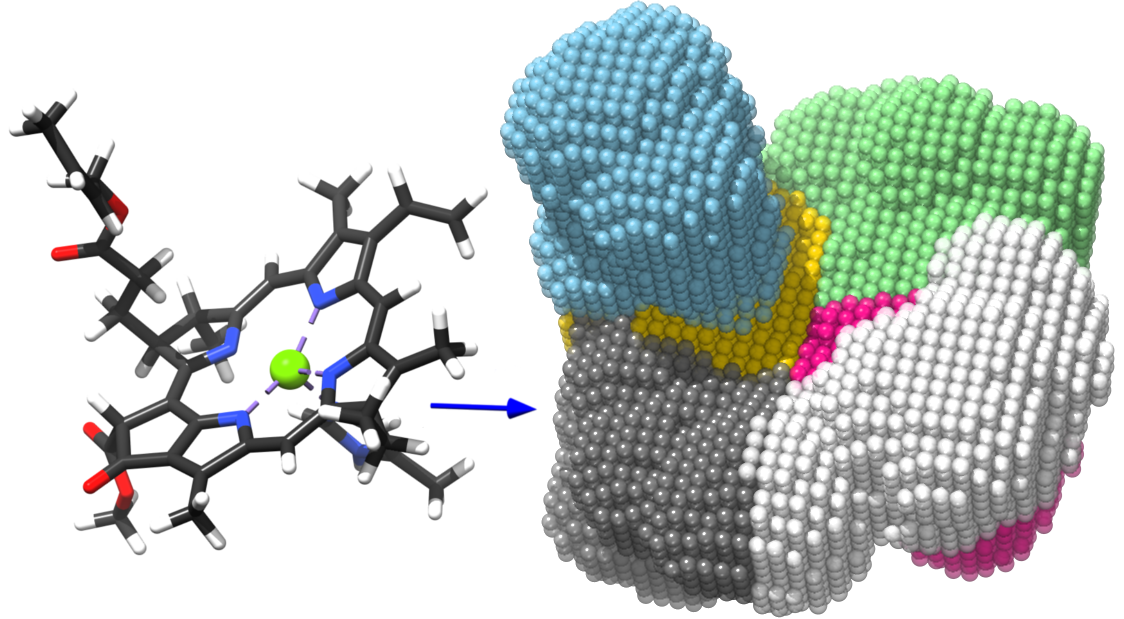}
\caption{\label{fig:partitioning}
Example of adaptive mesh partitioning for a molecule of chlorophyll~\textit{a}. Simplified mesh with a spacing of 0.5 \AA~ and a radius of 2.5 \AA. Each color represents a domain, which will be distributed to a set of processors for parallel computation.}
\end{center}
\end{figure}

Parallelization in Octopus is performed on different levels. The most
basic one is domain decomposition, were the grid is divided in
different regions that are assigned to each processor. For most
operations, only the boundaries of the regions need to be communicated
among processors. Since the grid can have a complicated shape dictated
by the shape of the molecule, it is far from trivial to distribute the grid-points among processors. For this task we use a third-party library called {\sc
  ParMETIS}~\cite{Karypis_1996}. This library provides routines to
partition the grid ensuring a balance of points and minimizing the size
of the boundary regions, and hence the communication costs. An example
of grid partitioning is shown in Fig.~\ref{fig:partitioning}.

Additional parallelization is provided by other data decomposition
approaches that are combined with domain decomposition. This includes parallelization over KS states, and over \(k\)-points and spin. The latter parallelization strategy is quite efficient, since for each \(k\)-point or spin component the operations are independent. However, it is limited by the size of the system, and often is not available (as in the case of closed-shell molecules, for example).

The efficiency of the parallelization over KS states depends on the type of calculation being performed. For ground state calculations, the orthogonalization and subspace diagonalization routines~\cite{Kresse_1996} require the communication of states. In Octopus this is handled by parallel dense linear-algebra operations provided by the ScaLAPACK library~\cite{scalapack}. For real-time propagation, on the other hand, the orthogonalization is preserved by the propagation~\cite{Castro_2006} and there is no need to communicate KS states between processors. This makes real-time TDDFT extremely efficient in massively parallel computers~\cite{Andrade_2012,Schleife_2014}.

An operation that needs special care in parallel is the solution of the
Poisson equation. Otherwise, it constitutes a bottleneck in parallelization, as a
single Poisson solution is required independently of the number of states in the system. A considerable effort has been devoted to the
problem of finding efficient parallel Poisson solvers that can keep up
with the rest of the code~\cite{Garc_a_Risue_o_2013}. We have found that the  most efficient methods are based on FFTs, which require a different domain
decomposition to perform efficiently. This introduces the additional
problem of transferring the data between the two different data
partitions. In Octopus this was overcome by creating a mapping at
the initialization stage and using it during execution to efficiently
communicate only the data that is strictly necessary between
processors~\cite{Alberdi_2014}.

\section{Conclusions}

In this article, we have shown several recent developments in the realm of electronic-structure theory that have been based on the Octopus real-space code and made possible in part by the flexibility and simplicity of working with real-space grids. Most of them go beyond a mere implementation of existing theory and represent new ideas in their respective areas. We expect that many of these approaches will become part of the standard tools of physicists, chemists and material scientists, and in the future will be integrated into other electronic-structure codes.

These advances also illustrate the variety of applications of real-space electronic structure, many of which going beyond the traditional calculation schemes used in electronic structure, and might provide a way forward to tackle current and future challenges in the field. 

What we have presented also shows some of the current challenges in real-space electronic structure. One example is the use of pseudo-potentials or other forms of projectors to represent the electron-ion interaction. Non-local potentials introduce additional complications on both the formulation, as shown by the case of magnetic response, and the implementation. Pseudo-potentials also include an additional, and in some cases, not well-controlled approximation. It would be interesting to study the possibility of developing an efficient method to perform full-potential calculations without additional computational cost, for example by using adaptive or radial grids.

Another challenge for real-space approaches is the cost of the calculation of two-body Coulomb integrals that appear in electron-hole linear response, RDMFT or hybrid xc functionals. In real-space these integrals are calculated in linear or quasi-linear time by considering them as a Poisson problem. However, the actual numerical cost can be quite large when compared with other operations. A fast approach to compute these integrals, perhaps by using an auxiliary basis, would certainly make the real-space approach more competitive for some applications.

The scalability of real-space grid methods makes them a good candidate for electronic-structure simulations in the future exaflop supercomputing systems expected for the end of the decade. In this aspect, the challenge is to develop high-performance implementations that can run efficiently on these machines.

%In general the number of grid points is large (in the range of \(10^4\) to \(10^6\)) in comparison with the number of expansion coefficients in localized basis set method. This is usually not an issue, since for real-space DFT/TDDFT the amount of work per coefficient is small and scales linearly with the number of grid points. However, some other methods are formulated in terms of objects that depend on two, or more, coordinates. For these systems, real-space methods become impractical even for moderately sized systems. In this case, alternative formulations are required that avoid these expensive many-body objects. This idea has been applied, for example, for the calculation of response functions~\cite{Nguyen_2012}.

\section{Acknowledgments}

We would like to thank all the people that have contributed to Octopus and to the development and implementation of the applications presented in this article. In particular we would like to acknowledge Silvana Botti, Jacob Sanders, Johanna Fuks, Heiko Appel, Danilo Nitsche, and Daniele Varsano.

XA acknowledges that part of this work was performed under the auspices of the U.S. Department of Energy at Lawrence Livermore National Laboratory under Contract DE-AC52-07A27344. XA and AA-G would like to thank the support received from Nvidia Corporation through the CUDA Center of Excellence program and the US Defense Threat Reduction Agency under contract no. HDTRA1-10-1-0046. DAS acknowledges support from the U.S. National Science Foundation graduate research program and IGERT fellowships, and from ARPA-E under grant DE-AR0000180. MJTO acknowledges financial support from the Belgian FNRS through FRFC project number 2.4545.12 ``Control of attosecond dynamics: applications to molecular reactivity''.
NH and IT received support from a Emmy-Noether grant from Deutsche Forschungsgemeinschaft.
JAR, AV, UDG, AHL and AR ackowledge financial support by the European Research Council Advanced Grant DYNamo (ERC-2010- AdG-267374), European Commission project CRONOS (Grant number 280879-2 CRONOS CP-FP7), Marie Curie ITN POCAONTAS (FP7-PEOPLE-2012-ITN, project number 316633), OST Actions CM1204 (XLIC), and MP1306 (EUSpec), Spanish Grant (FIS2013-46159-C3-1-P) and Grupo Consolidado UPV/EHU del Gobierno Vasco (IT578-13). JAR acknowledges the Department of Education, Universities and Research of the Basque Government (grant IT395-10).

\bibliography{converted_to_latex.bib}

\end{document}